\documentclass[aps,prl,reprint,twocolumn,superscriptaddress]{revtex4-1}
\usepackage{amsmath}
\usepackage{amsfonts}
\usepackage{amssymb}
\usepackage{amsthm}
\usepackage{graphicx}
\usepackage{CJK}
\usepackage{times}
\usepackage{mathptmx}
\usepackage[colorlinks, linkcolor=blue, urlcolor=blue, anchorcolor=blue, citecolor=blue]{hyperref}
\usepackage{mathtools}
\usepackage{lipsum}
\usepackage{array}
\usepackage[table]{xcolor}
\definecolor{mygray}{gray}{.9}

\begin{document}


\title{Ground states and magnonics in orthogonally-coupled symmetric all-antiferromagnetic junctions}



\author{Mei Li}
\affiliation{College of Physics Science and Technology, Yangzhou University, Yangzhou 225002, People's Republic of China}
\author{Bin Xi}
\affiliation{College of Physics Science and Technology, Yangzhou University, Yangzhou 225002, People's Republic of China}
\author{Wei He}
\affiliation{State Key Laboratory of Magnetism and Beijing National Laboratory for Condensed Matter Physics, Institute of Physics, Chinese Academy of Sciences, Beijing 100190, People's Republic of China}
\author{Yongjun Liu}
\affiliation{College of Physics Science and Technology, Yangzhou University, Yangzhou 225002, People's Republic of China}
\author{Jie Lu}
\email{lujie@yzu.edu.cn}
\affiliation{College of Physics Science and Technology, Yangzhou University, Yangzhou 225002, People's Republic of China}


\date{\today}

\begin{abstract}
In this work, the rich ground-state structure of orthogonally-coupled symmetric
all-antiferromagnetic junctions with easy-plane anisotropy is reported.
Spin reorientation process rather than the traditional spin flop (SF) occurs,
resulting in a novel phase in which N\'{e}el vectors preserve the mirror-reflection symmetry (termed as ``MRS phase").
The phase transitions between SF and MRS phases  can be either the first- or second-order.
After disturbed by external stimuli, magnons with different parities emerge. 
For in-plane dc fields, no couplings between magnons occur.
When dc fields become oblique, coherent couplings between magnons with opposite parity emerge,
leading to anticrossings in resonance frequencies. 
However, self-hybridization among magnons with the same parity never happens.
More interestingly, spin waves based on MRS phase are linearly polarized and their 
polarization directions can be fine controlled.
\end{abstract}


\maketitle


\section{I. Introduction} 
All-antiferromagnetic junctions, 
such as $\mathrm{Fe}_2\mathrm{O}_3$/$\mathrm{Cr}_2\mathrm{O}_3$/$\mathrm{Fe}_2\mathrm{O}_3$, 
have been recently proposed to be candidates for practical antiferromagnetic spintronics and magnonics with 
ultrahigh-density integration\cite{Zhou_NC_2022}.
Different from the well-know synthetic antiferromagnets\cite{Duine_nphys_2018,Martin_JAP_2007,Belmeguenai_JPCM_2008,Seki_APL_2009,Gonzalez_PRB_2013,Timopheev_PRB_2014,Liu_PRB_2014,Tanaka_APExpress_2014,Yang_APL_2016,Li_AFM_2016,CWang_APL_2018,WWang_APL_2018,XFHan_APL_2018,Kamimaki_APL_2019,Chen_APL_2019,Sorokin_PRB_2020,Kamimaki_PRAppl_2020,Shiota_PRL_2020,Shiota_SciAdv_2020,Sud_PRB_2020,Waring_PRApplied_2020,LiMei_PRB_2021,HeWei_CPL_2021,Troncoso_PRB_2021,Ma_APL_2021,Chen_JPCM_2022} 
in which the collinear interlayer coupling between
ferromagnetic sublayers is the 
Ruderman-Kittel-Kasuya-Yosida (RKKY) interaction mediated by electrons\cite{Kittel_PhyRev_1954,Yafet_JAP_1987,Bruno_PRL_1991},
the coupling between two antiferromagnetic $\mathrm{Fe}_2\mathrm{O}_3$ sublayers in 
all-antiferromagnetic junctions is claimed to be bridged by the non-uniform domain wall state 
in $\mathrm{Cr}_2\mathrm{O}_3$ spacers.
In recent experiments, the  double-peak structure in spin Hall magnetoresistance (SMR)\cite{Nakayama_PRL_2013,Chen_PRB_2013,Hoogeboom_APL_2017,Cheng_PRB_2019,Fischer_PRApplied_2020}
signals of asymmetric $\mathrm{Fe}_2\mathrm{O}_3$/$\mathrm{Cr}_2\mathrm{O}_3$/$\mathrm{Fe}_2\mathrm{O}_3$  junctions
are explained by the orthogonal coupling between the N\'{e}el vectors in the two outermost
$\mathrm{Fe}_2\mathrm{O}_3$ layers with different thickness.
However, the detailed magnetic ground states of these junctions have not yet been provided.
In addition, magnon-magnon coupling and its resulting modification to the eigenfrequencies
and eigenvectors need to be revealed. 
In this work, we focus on the simplified version, that is, symmetric all-antiferromagnetic junctions
where the two outermost sublayers are made from the same materials with the same thickness.
The magnetic ground states and coherent magnonics in these systems constitute the main contents of this work.

\section{II. Model and methods} 
\begin{figure} [htbp]
	\centering
	\includegraphics[width=0.35\textwidth]{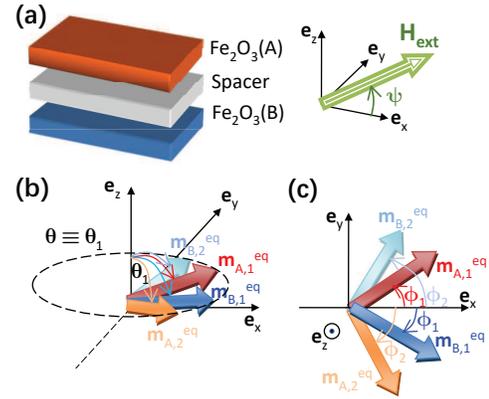}
	\caption{(Color online) (a) Sketch of an orthogonally-coupled symmetric all-antiferromagnetic junction
		with saturation magnetization $M_s$	and thickness $d$ in two outermost $\mathrm{Fe}_2\mathrm{O}_3$ sublayers. 
		$\mathbf{e}_{z}$ is along surface normal, $\mathbf{e}_y \parallel \mathbf{e}_z\times \mathbf{H}_{\mathrm{ext}}$, 
		and $\mathbf{e}_x=\mathbf{e}_y \times \mathbf{e}_z $, in which the external oblique dc field reads
		$\mathbf{H}_{\mathrm{ext}}=H_0(\cos\psi\mathbf{e}_x+\sin\psi\mathbf{e}_z)$.
		(b) The four equilibrium unit magnetization vectors in two $\mathrm{Fe}_2\mathrm{O}_3$ sulayers are denoted as
		$\mathbf{m}_{A(B),1(2)}^{\mathrm{eq}}$ and all fall onto the same ``latitude circle" $\theta=\theta_1$.
		Meantime, $\mathbf{m}_{A,1}^{\mathrm{eq}}$ and $\mathbf{m}_{B,1}^{\mathrm{eq}}$ 
		($\mathbf{m}_{A,2}^{\mathrm{eq}}$ and $\mathbf{m}_{B,2}^{\mathrm{eq}}$) are symmetric about the $xz-$plane.
		(c) Vertical view of (b).
		The azimuthal angle of $\mathbf{m}_{A,1}^{\mathrm{eq}}$ ($\mathbf{m}_{B,1}^{\mathrm{eq}}$) is $\phi_1$ ($-\phi_1$),
		while that for $\mathbf{m}_{A,2}^{\mathrm{eq}}$ ($\mathbf{m}_{B,2}^{\mathrm{eq}}$) is $-\phi_2$ ($\phi_2$).
	}\label{fig1}
\end{figure}

Generally the antiferromagnetic $\mathrm{Cr}_2\mathrm{O}_3$ spacer possesses a spin-flop (SF) field\cite{Anderson_PR_1964}
higher than 6 T\cite{Dayhoff_PR_1957,Foner_PR_1963},
which is much larger than that of $\mathrm{Fe}_2\mathrm{O}_3$ (a few thousand Oe)\cite{Lebrun_NC_2020,Han_Nanotech_2020}.
Therefore in our analytics the two identical antiferromagnetic outermost $\mathrm{Fe}_2\mathrm{O}_3$ sublayers (denoted as ``A, B") 
are assumed to have the simplest ``easy-plane" anisotropy with the hard axis along the surface normal ($z$ axis), 
as shown in Fig. \ref{fig1}(a).
In each layer, the magnetic energy density is:
\begin{equation}\label{Single_AFM_energy_density}
\begin{split}
	\frac{\mathcal{E}_{\mathrm{s}}\left[\mathbf{m}_{i,1(2)}\right]}{\mu_0}  =& M_s H_E \mathbf{m}_{i,1}\cdot \mathbf{m}_{i,2}+\frac{M_s H_K}{2}\left(\mathrm{m}_{i,1z}^2+\mathrm{m}_{i,2z}^2\right)   \\
	& - M_s\left(\mathbf{m}_{i,1}+\mathbf{m}_{i,2}\right)\cdot \mathbf{H}_{\mathrm{ext}},\quad i=A,B,
\end{split}
\end{equation}
where $\mathbf{m}_{i,1}$ and $\mathbf{m}_{i,2}$ are the unit vectors of the two sublattices in the $i-$th antiferromagnetic sublayer
with the same saturation magnetization $M_s$ and mutual exchange field $H_E (>0)$, 
$H_K (>0)$ is the anisotropy field along hard axis, 
$\mathbf{H}_{\mathrm{ext}}$ denotes the external dc field, and $\mu_0$ is the vacuum permeability.

~\\
~\\

We then neglect the details of spacers and phenomenologically introduce the orthogonal coupling between two outermost sublayers. 
Originally, the interlayer orthogonal coupling is expressed as
$\mathcal{E}_{\mathrm{c}}=J'(\mathbf{n}_A\cdot\mathbf{n}_B)^2/2=J'\cos^2\theta_{AB}/2$, in which $J'>0$ and
$\mathbf{n}_i\equiv (\mathbf{m}_{i,1}- \mathbf{m}_{i,2})/|\mathbf{m}_{i,1}- \mathbf{m}_{i,2}|$
are the N\'{e}el vectors of each antiferromagnetic sublayer with $\theta_{AB}$ being the angle spanned by them.
However, this format of $\mathcal{E}_{\mathrm{c}}$ is inconvenient in deducing the equilibrium ground states and dynamic responses of
magnetization vectors.
Considering the fact that in antiferromagnets, $\mathbf{n}_i\perp\mathbf{m}_i (\equiv \mathbf{m}_{i,1} +\mathbf{m}_{i,2} $),  
an equivalent format is provided as
\begin{equation}\label{Coupling_energy_density}
	\mathcal{E}_{\mathrm{c}}=\frac{J}{2}\left(\mathbf{m}_A\cdot\mathbf{m}_B\right)^2,
\end{equation}
with $J(>0)$ denotes the coupling strength. 
Combing all these components, the total magnetic energy of this junction reads
\begin{equation}\label{Total_energy}
E_{\mathrm{tot}}=\sum_{i=A,B} \mathcal{E}_{\mathrm{s}}\left[\mathbf{m}_{i,1},\mathbf{m}_{i,2}\right] \cdot(S_0 d_0) + 
\mathcal{E}_{\mathrm{c}}\cdot S_0,
\end{equation}
with $S_0$ and $d_0$ being the projection area and thickness of each outermost sublayer, respectively.
In macrospin assumption, the effective field of a ``single-domain" volume with unit magnetization $\mathbf{m}$,
projection area $S$ and thickness $d$ is defined as 
$\mathbf{H}_{\mathrm{eff}}=-\mu_0^{-1}\delta E_{\mathrm{tot}}/\delta (M_s S d \mathbf{m})$.

The dynamics of $\mathbf{m}_{A(B),1(2)}$ are described by the coupled Landau-Lifshitz-Gilbert equations\cite{LLG_equation}
\begin{equation}\label{Coupled_LLG_vectorial}
\frac{d\mathbf{m}_{A(B),1(2)}}{dt}=-\gamma \mathbf{m}_{A(B),1(2)}\times \mathbf{H}_{\mathrm{eff}}^{A(B),1(2)} +\mathbf{T}_{A(B),1(2)},
\end{equation}
where $\mathbf{T}_{A(B),1(2)}$ includes torques from both damping and external dc fields, 
$\gamma=\mu_0\gamma_e$ with $\gamma_e$ being electron gyromagnetic ratio. 
The effective field $\mathbf{H}_{\mathrm{eff}}^{A,1}$ reads
\begin{equation}\label{H_eff_A1}
\mathbf{H}_{\mathrm{eff}}^{A,1}=\mathbf{H}_{\mathrm{ext}}-H_E\mathbf{m}_{A,2}-H_K m_{A,1z}\mathbf{e}_z-H_p \left(\mathbf{m}_A\cdot\mathbf{m}_B\right)\mathbf{m}_B,
\end{equation}
with $H_p\equiv J/(\mu_0 M_s d)$,
the other three effective fields can be obtained by performing $A\leftrightarrow B$ and $1\leftrightarrow 2$.
These are all we need to proceed our investigation.

\section{III. Ground states and phase transitions} 

\begin{figure} [htbp]
	\centering
	\includegraphics[width=0.45\textwidth]{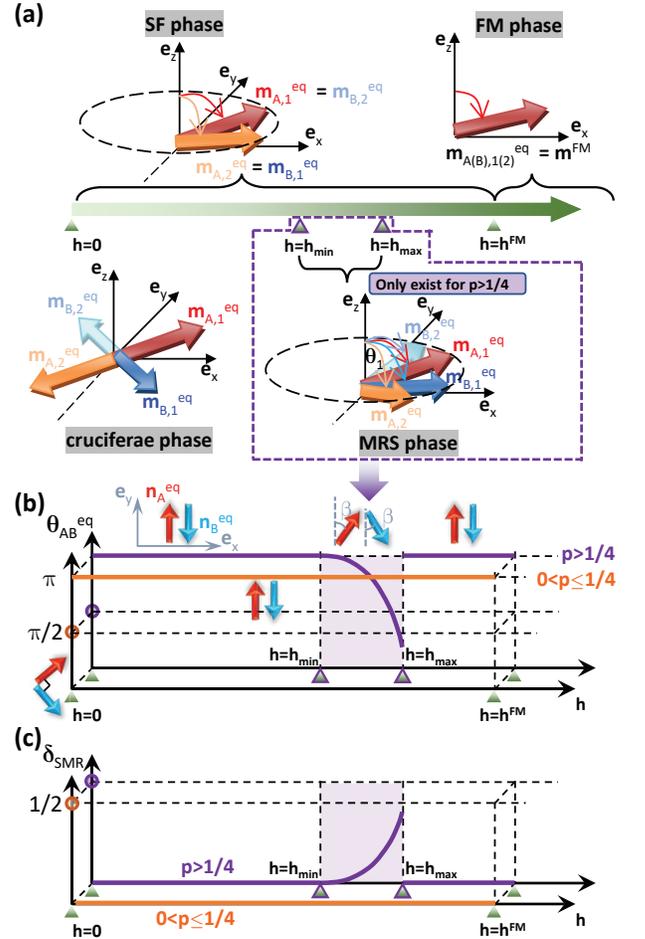}
	\caption{(Color online) (a) Ground-state evolution of orthogonally-coupled symmetric all-antiferromagnetic junctions
		as the external dc field increases. At $h=0$, $\mathbf{m}_{A(B),1(2)}$ evenly spread in $xy-$plane
		at $\pi/2$ interval (``cruciferae" phase). 
		For $h\ge h^{\mathrm{FM}}$, $\mathbf{m}_{A(B),1(2)}\equiv\mathbf{m}^{\mathrm{FM}}$
		thus the system falls into the FM phase.
		When $0<h<h^{\mathrm{FM}}$, the SF phase with $\mathbf{m}_{A,1}^{\mathrm{eq}}=\mathbf{m}_{B,2}^{\mathrm{eq}}$
		and $\mathbf{m}_{A,2}^{\mathrm{eq}}=\mathbf{m}_{B,1}^{\mathrm{eq}}$ bears the lowest magnetic energy for $0<p<1/4$. 
		Only for $p>1/4$, a MRS phase emerges and acquires the lowest energy
		for $h\in[h_{\mathrm{min}},h_{\mathrm{max}}]\subset (0,h^{\mathrm{FM}})$.
		(b) Variation of $\theta_{AB}^{\mathrm{eq}}$ (the angle between $\mathbf{n}_A^{\mathrm{eq}}$ 
		and $\mathbf{n}_B^{\mathrm{eq}}$) with the external fields. 
		Orange line (purple curve) is for the case with $0<p<1/4$ ($p>1/4$). 
		Red (blue) arrow denotes $\mathbf{n}_A^{\mathrm{eq}}$ ($\mathbf{n}_B^{\mathrm{eq}}$) in various phases.
		The PT (SF process) at $h=0$ is always the first-order as $\theta_{AB}^{\mathrm{eq}}$ changes
		abruptly from $\pi/2$ to $\pi$. 
		For $p>1/4$, the light purple area denotes the novel MRS phase. 
		The PT between SF and MRS phases at $h=h_{\mathrm{min}}$ ($h_{\mathrm{max}}$) is the second (first) order.
		(c) Normalized SMR signals as functions of the external field $h$. Orange line (purple curve) is for the 
		case with $0<p<1/4$ ($p>1/4$). 
	}\label{fig2}
\end{figure}

We consider a general $\mathbf{H}_{\mathrm{ext}}$
with a strength $H_0$ and polar angle $\frac{\pi}{2}-\psi$ with respect to $\mathbf{e}_z$ [see Fig. \ref{fig1}(a)]. 
After defining $\mathbf{e}_y \parallel \mathbf{e}_z\times \mathbf{H}_{\mathrm{ext}}$ and $\mathbf{e}_x=\mathbf{e}_y \times \mathbf{e}_z $,
$\mathbf{H}_{\mathrm{ext}}=H_0(\cos\psi\mathbf{e}_x+\sin\psi\mathbf{e}_z)$.
The equilibrium magnetization vectors (emphasized by a superscript ``eq") in sublayers A and B should 
be symmetric about $xz-$plane.
In view of this, we define the following polar and azimuthal angles: 
$\mathbf{m}_{A,1(2)}^{\mathrm{eq}}$ ($\mathbf{m}_{B,1(2)}^{\mathrm{eq}}$) has the polar angle $\theta_{1(2)}$ and
azimuthal angle $\phi_{1(2)}$ ($-\phi_{1(2)}$).
The static condition requires $d\mathbf{m}_{A(B),1(2)}/dt=0$ and $\mathbf{T}_{A(B),1(2)}=0$.
This results in: 
(i) $\theta_1=\theta_2$, that is, $\mathbf{m}_{A(B),1(2)}$ all lie in the same ``latitude circle" [see Fig. \ref{fig1}(b)], 
hence $\mathbf{n}_{A}^{\mathrm{eq}}$ and $\mathbf{n}_B^{\mathrm{eq}}$ always reside in the $xy-$plane;
(ii) the central equality of this work
\begin{equation}\label{Equality}
	h(\cos\psi\mathbf{e}_x+\sin\psi\mathbf{e}_z)=\mathbf{m}_{A}^{\mathrm{eq}}+p\left(\mathbf{m}_{A}^{\mathrm{eq}}\cdot\mathbf{m}_{B}^{\mathrm{eq}}\right)\mathbf{m}_{B}^{\mathrm{eq}}+k\cos\theta_1\mathbf{e}_z,
\end{equation}
where $h\equiv H_0/H_E$, $p\equiv H_p/H_E$, and $k\equiv H_K/H_E$, which are all positive.
This equality is very important and will be used to: 
(i) calculate $\mathbf{m}_{A(B),1(2)}^{\mathrm{eq}}$ (i.e. $\theta_1$ and $\phi_{1(2)}$);
(2) simplify the dynamic equations for spin waves.

First two extreme circumstances are examined: zero-field and high-enough-field cases.
In the absence of external fields ($h=0$), due to the lack of in-plane anisotropy
the ground state of this symmetric all-antiferromagnetic junction 
is a ``cruciferae" state, namely $\mathbf{m}_{A,1}$, $\mathbf{m}_{B,1}$, $\mathbf{m}_{A,2}$, and $\mathbf{m}_{B,2}$
are successively arranged clockwise in $xy-$plane at $\pi/2$ intervals (thus $\theta_{AB}^{\mathrm{eq}}=\pi/2$),
taking the most advantage of the energy loss from the orthogonal coupling.
On the other hand, there always exists an upper limit of external field strength
$h^{\mathrm{FM}}\equiv\left[\cos^2\psi/(8p+2)^2 + \sin^2\psi/(8p+k+2)^2\right]^{-1/2}$.
Above it, the system falls into the ferromagnetic saturated state (FM phase), which means 
$\mathbf{m}_{A(B),1(2)}\equiv\mathbf{m}^{\mathrm{FM}}=\sin\theta_1^{\mathrm{FM}}\mathbf{e}_x+\cos\theta_1^{\mathrm{FM}}\mathbf{e}_z$ with $\theta_1^{\mathrm{FM}}$ satisfying
\begin{equation}\label{Theta_1_FM}
	h\cos\left(\theta_1^{\mathrm{FM}}+\psi\right)+(k/2)\sin2\theta_1^{\mathrm{FM}}=0.
\end{equation}
The presence of $k$ leads to $\pi/2-\psi<\theta_1^{\mathrm{FM}}<\pi/2$, meaning that $\mathbf{m}^{\mathrm{FM}}$
lies between $\mathbf{H}_{\mathrm{ext}}$ and $\mathbf{e}_x$.
Now $\mathbf{n}_A^{\mathrm{eq}}=\mathbf{n}_B^{\mathrm{eq}}=0$, hence $\theta_{AB}$ has no definition in FM phase.

For $0<h<h^{\mathrm{FM}}$, the system exhibits complex ground states and phase transitions between them,
as illustrated in Figs. \ref{fig2}(a) and \ref{fig2}(b).
The orthogonal coupling strength $p$ strongly affect the equilibrium magnetization layout.
For weak orthogonal couplings ($0<p<1/4$), before saturation the system always falls into 
the SF ground state with the polar and azimuthal angles satisfying
\begin{equation}\label{Symmetrical_Flower_theta1_phi1}
	\begin{split}
		&  8pk^2\cos^5\theta_1-16phk\sin\psi\cos^4\theta_1   \\
		&   +(8ph^2+k^3+2k^2)\cos^3\theta_1-(3k+4)hk\sin\psi\cos^2\theta_1   \\
		&   +(3k+2)h^2\sin^2\psi\cos\theta_1-h^3\sin^3\psi=0,  \\
		& \cos\phi_1=\cos\phi_2=\frac{h\cos\psi\cos\theta_1}{\left(h\sin\psi-k\cos\theta_1\right)\sin\theta_1}.
	\end{split}
\end{equation}
In fact, this SF phase is the counterpart of spin-flopped state in a single antiferromagnetic layer (with zero in-plane anisotropy)
since the orthogonal coupling now is too weak to affect the magnetization layout.
Now $\mathbf{n}_A^{\mathrm{eq}}=\mathbf{e}_y$ and $\mathbf{n}_B^{\mathrm{eq}}=-\mathbf{e}_y$,
leading to $\theta_{AB}^{\mathrm{eq}}=\pi$.
Correspondingly, the phase transition (PT) at $h=0$ is the first-order.
At $h=h^{\mathrm{FM}}$, Eq. (\ref{Symmetrical_Flower_theta1_phi1}) continuously converges
to Eq. (\ref{Theta_1_FM}), leading to a second-order PT therein.

More interestingly, for large enough orthogonal coupling ($p>1/4$), 
a new phase acquires lower energy than the SF phase
for $h\in[h_{\mathrm{min}},h_{\mathrm{max}}]\subset (0,h_{\mathrm{max}}^{\mathrm{up}})$ with
$h_{\mathrm{min}}\equiv(p^{-1/2}/2)/\sqrt{\cos^2\psi/16+\sin^2\psi/(k+4)^2}$ and
$h_{\mathrm{max}}\equiv\sqrt{1/2+1/(8p)}/\sqrt{\cos^2\psi/16+\sin^2\psi/(k+4)^2}$.
The polar and azimuthal angles are
\begin{equation}\label{Asymmetrical_Flower_theta1_phi1_phi2}
	\begin{split}
       \cos\theta_1=&\frac{h}{k+4}\sin\psi ,\quad \phi_1=\alpha-\beta,\quad \phi_2=\alpha+\beta, \\
       \mathrm{with}\quad \alpha\equiv& \arccos\frac{1}{2}\sqrt{\frac{\frac{\cos^2\psi}{2}+\frac{4\sin^2\psi}{(k+4)^2}-\frac{1}{h^2 p}}{\frac{1}{h^2}-\frac{\sin^2\psi}{(k+4)^2}}},  \\
       \beta\equiv & \arccos\sqrt{\frac{\frac{\cos^2\psi}{4}}{\frac{\cos^2\psi}{2}+\frac{4\sin^2\psi}{(k+4)^2}-\frac{1}{h^2 p}}},
	\end{split}
\end{equation}
where we have assumed $|\phi_1|<\phi_2<\pi-|\phi_1|$ without losing generality. 
Then $\mathbf{n}_{A}^{\mathrm{eq}}=\sin\beta\mathbf{e}_x+\cos\beta\mathbf{e}_y$
and $\mathbf{n}_{B}^{\mathrm{eq}}=\sin\beta\mathbf{e}_x-\cos\beta\mathbf{e}_y$.
Correspondingly, $\theta_{AB}^{\mathrm{eq}}=\pi-2\beta=\pi-(\phi_2-\phi_1)$.
This phase comes from the competition between interlayer orthogonal and intralayer antiferromagnetic couplings,
however it preserves the mirror-reflection symmetry (MRS) about the  $xz-$plane.
We hereby denote it as the ``MRS phase".
It coincides with the SF phase at $h=h_{\mathrm{min}}$ since $\beta(h_{\mathrm{min}})=0$ (hence $\phi_1=\phi_2$),
which implies a second-order PT herein ($\theta_{AB}^{\mathrm{eq}}=\pi$ on both sides of $h_{\mathrm{min}}$).
However at $h=h_{\mathrm{max}}$,  $\alpha=0$ hence $-\phi_1=\phi_2=\beta(h_{\mathrm{max}})$
with $\beta(h_{\mathrm{max}})\equiv\arccos\sqrt{(4p+1)/[8p+16(4p-1)\tan^2\psi/(k+4)^2]}<\pi/2$.
Now $\mathbf{m}_{A,1}=\mathbf{m}_{A,2}$ and $\mathbf{m}_{B,1}=\mathbf{m}_{B,2}$,
leading to $\theta_{AB}^{\mathrm{eq}}=\pi-2\beta(h_{\mathrm{max}})>0$.
When $h$ slightly exceeds $h_{\mathrm{max}}$, the ground state abruptly moves back
to the SF phase with $\theta_{AB}^{\mathrm{eq}}=\pi$, leading to a first-order PT.

This novel MRS phase is essential for the non-zero SMR signals under
finite dc fields even in the absence of in-plane magnetic anisotropy for symmetric all-antiferromagnetic junctions
with strong enough orthogonal couplings.
Suppose a heavy-metal caplayer (for example, Pt), through which a charge current passes in $+\mathbf{e}_x$ direction,
is attached onto a symmetric all-antiferromagnetic junction under $\mathbf{H}_{\mathrm{ext}}=hH_E(\cos\psi\mathbf{e}_x+\sin\psi\mathbf{e}_z)$.
The spin polarization generated by the spin Hall effect of the heavy-metal caplayer is then along the $y-$axis.
Generally, when the N\'{e}el vectors of the outermost antiferromagnetic sublayers 
($\mathbf{n}_{A}^{\mathrm{eq}}$ and $\mathbf{n}_{B}^{\mathrm{eq}}$)
are parallel (perpendicular) to the spin polarization, 
the system has a comparatively low (high) resistance.
In general,  we use
\begin{equation}\label{Normalized_SMR}
	\delta_{\mathrm{SMR}}=1-\frac{\left(\mathbf{n}_{A}^{\mathrm{eq}}\cdot\mathbf{e}_y\right)^2+\left(\mathbf{n}_{B}^{\mathrm{eq}}\cdot\mathbf{e}_y\right)^2}{2}
\end{equation}
to describe the normalized SMR signal of this junction, as illustrated in Fig. \ref{fig2}(c).

For small orthogonal coupling ($0<p<1/4$), $\mathbf{n}_{A(B)}^{\mathrm{eq}}$ in SF 
phase always lie in $y-$axis for nonzero $h$,
leading to a plateau with $\delta_{\mathrm{SMR}}^{\mathrm{SF}}=0$ [see orange line in Fig. \ref{fig2}(c)]. 
The only peak with the height of $1/2$ occurs at $h=0$ where the cruciferae phase emerges.
While for large enough orthogonal coupling ($p>1/4$), in the MRS phase one has
$\delta_{\mathrm{SMR}}^{\mathrm{MRS}}=\sin^2\beta$
which is strengthened from 0 to 
\begin{equation}\label{Normalized_SMR_MRSphase}
	\delta_{\mathrm{SMR}}^{\mathrm{MRS}}(\beta_{\mathrm{max}})=\frac{1+\frac{16\tan^2\psi}{(k+4)^2}}{\frac{8p}{4p-1}+\frac{16\tan^2\psi}{(k+4)^2}}
\end{equation}
as $h$ increases from $h_{\mathrm{min}}$ to $h_{\mathrm{max}}$.
The normalized SMR signal rapidly decreases to zero as $h$ exceeds $h_{\mathrm{max}}$.
In real junctions, these abrupt peaks will be broadened due to the finite distribution of the anisotropy in sublayers, 
as well as the possible magnetic impurities therein. 

This MRS-phase-induced resistance peak is unique for symmetric all-antiferromagnetic junctions since 
$\mathbf{n}_{A(B)}^{\mathrm{eq}}$ lie symmetrically about $xz-$plane.
It differs from the resistance peaks (i) and (iii) in Figs. 2(d) and 2(e) from Ref. \cite{Zhou_NC_2022}, 
at which $\mathbf{n}_{A}^{\mathrm{eq}}$ and $\mathbf{n}_{B}^{\mathrm{eq}}$
are respectively along $\mathbf{e}_x$ and $\mathbf{e}_y$. 
In addition, it discriminates from the negative SMR signal of a single antiferromagnetic
$\mathrm{Fe}_2\mathrm{O}_3$ layer
(see Figs. 2(a) and 2(b) from Ref. \cite{Zhou_NC_2022}), where the (only) resistance peak 
appears at $H<0$ ($H>0$) when in-plane external field with strength $H$ sweeping from 
positive to negative (negative to positive). 
The reasons are twofold. 
On one hand, the MRS-phase-induced resistance peak survives
even in the absence of in-plane anisotropy while that of single $\mathrm{Fe}_2\mathrm{O}_3$ layer relies on it.
On the other hand, the MRS peak occurs in the MRS phase which emerges for 
$h\in[h_{\mathrm{min}},h_{\mathrm{max}}]\subset(0,h^{\mathrm{FM}})$ thus is a ``post-SF" effect
taking place under a relatively large external field strength (around $h_{\mathrm{max}}$).
The latter reason also helps to exclude the interlayer coupling between $\mathrm{Fe}_2\mathrm{O}_3$ sublayers
in $\mathrm{Fe}_2\mathrm{O}_3$/$\mathrm{Al}_2\mathrm{O}_3$/$\mathrm{Fe}_2\mathrm{O}_3$ junctions
as shown in Fig. S19 from the Supplementary Note of Ref. \cite{Zhou_NC_2022}.
At last, the experimental SMR data of symmetric $\mathrm{Fe}_2\mathrm{O}_3$/$\mathrm{NiO}$/$\mathrm{Fe}_2\mathrm{O}_3$
junction in Fig. S18 from the same work did not show strong signs of MRS peak.
This possibly comes from the fact that NiO has not-high-enough SF field\cite{JPCSSP_1980_NiO_SF_1,JPCSSP_1980_NiO_SF_2}
thus makes the data contain too many physical processes.

In the end of this section, we define several symmetry operators for further usage . 
The first one is the mirror reflection operator $\mathbb{M}_{xz}$
which converts a vector $\mathbf{U}=U_x\mathbf{e}_x+U_y\mathbf{e}_y+U_z\mathbf{e}_z$
to $\mathbb{M}_{xz}\mathbf{U}=U_x\mathbf{e}_x-U_y\mathbf{e}_y+U_z\mathbf{e}_z$.
$\mathbb{M}_{xz}$ is used to simplify the dynamical equations of spin waves in MRS phase 
under oblique dc fields (to be delivered in Sec. IV.C.1),
but the polarization of spin waves will be reversed therefore we seldom use it unless necessary.
Next, we define two rotational operations which preserve the polarization of spin waves.
Since $\mathbb{M}_{xz}\mathbf{m}_{A,1(2)}^{\mathrm{eq}}=\mathbf{m}_{B,1(2)}^{\mathrm{eq}}$,
the planes expanded by $(\mathbf{m}_{A,1}^{\mathrm{eq}},\mathbf{m}_{B,1}^{\mathrm{eq}})$ 
and $(\mathbf{m}_{A,2}^{\mathrm{eq}},\mathbf{m}_{B,2}^{\mathrm{eq}})$
both intersect with $xy-$plane on the $y-$axis.
We choose $\mathbf{e}_{x^{\prime}}\parallel \mathbf{m}_{A,1}^{\mathrm{eq}}+\mathbf{m}_{B,1}^{\mathrm{eq}}$, 
and $\mathbf{e}_{x^{\prime\prime}}\parallel \mathbf{m}_{A,2}^{\mathrm{eq}}+\mathbf{m}_{B,2}^{\mathrm{eq}}$.
Correspondingly, the angle between $\mathbf{e}_{x^{\prime}}$ ($\mathbf{e}_{x^{\prime\prime}}$) 
and $\mathbf{e}_x$ is defined as $\chi$ ($\xi$). 
Meantime, the angle between $\mathbf{e}_{x^{\prime}}$ and $\mathbf{m}_{A,1}^{\mathrm{eq}}$ 
($\mathbf{e}_{x^{\prime\prime}}$ and $\mathbf{m}_{A,2}^{\mathrm{eq}}$) is $\epsilon$ ($\eta$).
Then we have $\tan\chi=\cot\theta_1/\cos\phi_1$, $\tan\xi=\cot\theta_1/\cos\phi_2$, 
$\sin\epsilon=\sin\theta_1\sin\phi_1$, and $\sin\eta=\sin\theta_1\sin\phi_2$.
Consequently, $\mathbb{C}_{2x^{\prime}}$ ($\mathbb{C}_{2x^{\prime\prime}}$) is defined as the 
rotation operator which rotates vectors around $+\mathbf{e}_{x^{\prime}}$ ($+\mathbf{e}_{x^{\prime\prime}}$) by $180^{\circ}$,
meantime preserves the polarization of spin waves.
With these ground states and operators, in the next sections we proceed to dynamical response of magnetization 
vectors to external stimuli.

\section{IV. Magnonics} 
\subsection{IV.A General framework}
Suppose a RF magnetic field with a frequency $f=\omega/2\pi$ is induced by some antenna
and then excites spin waves to propagate in the whole junction.
The magnetization vectors $\mathbf{m}_{A(B),1(2)}$ slightly deviate from
their equilibrium orientations $\mathbf{m}_{A(B),1(2)}^{\mathrm{eq}}$ and begin to vibrate with the same frequency $f$.
Consequently, they are expanded as:  
$\mathbf{m}_{A(B),1(2)}=\mathbf{m}_{A(B),1(2)}^{\mathrm{eq}}+\delta\mathbf{m}_{A(B),1(2)} e^{\mathrm{i}\omega t}$.
After defining $\Omega\equiv \omega/(\gamma H_E)$ and neglecting $\mathbf{T}_{A(B),1(2)}$, we obtain
the following central vectorial equations for $\delta\mathbf{m}_{A(B),1(2)}$ with the help of Eq. (\ref{Equality}),
\begin{widetext} 
	\begin{equation}\label{Coupled_LLG_most_general}
		\begin{split}
			\mathrm{i}\Omega\delta\mathbf{m}_{A(B),1(2)}=&\mathbf{m}_{A(B),1(2)}^{\mathrm{eq}}\times\left\{\left(\delta\mathbf{m}_{A(B),1}+\delta\mathbf{m}_{A(B),2}\right)+k(\delta\mathbf{m}_{A(B),1(2)}\cdot\mathbf{e}_z)\mathbf{e}_z
			+p\left(\mathbf{m}_A^{\mathrm{eq}}\cdot\mathbf{m}_B^{\mathrm{eq}}\right)\left(\delta\mathbf{m}_{B(A),1}+\delta\mathbf{m}_{B(A),2}\right)\right.  \\ 
			&\qquad\qquad\qquad
    		 \left.+p\left[\left(\delta\mathbf{m}_{A,1}+\delta\mathbf{m}_{A,2}\right)\cdot\mathbf{m}_B^{\mathrm{eq}}+
			 \left(\delta\mathbf{m}_{B,1}+\delta\mathbf{m}_{B,2}\right)\cdot\mathbf{m}_A^{\mathrm{eq}}\right]\mathbf{m}_{B(A)}^{\mathrm{eq}}\right\},
		\end{split}
	\end{equation}
\end{widetext} 
which are the central equations for coherent magnonics in orthogonally-coupled symmetric all-antiferromagnetic junctions.

\subsection{IV.B In-plane-dc-field case}
\subsubsection{IV.B.1 Reformation of dynamical equations}
For in-plane dc fields ($\psi=0$), $\theta_1=\pi/2$ thus $\mathbf{m}_{A(B),1(2)}^{\mathrm{eq}}$ all reside in $xy-$plane.
Consequently, $\mathbf{e}_{x^{\prime}}=\mathbf{e}_{x^{\prime\prime}}=\mathbf{e}_{x}$,
leading to $\mathbb{C}_{2 x^{\prime}}=\mathbb{C}_{2 x^{\prime\prime}}=\mathbb{C}_{2 x}$.
We then introduce 
$\delta\mathbf{m}_{1(2)}^{\pm}\equiv\delta\mathbf{m}_{A,1(2)} \pm \mathbb{C}_{2x}\delta\mathbf{m}_{B,1(2)}$ 
as the spin wave components with even ($+$) and odd ($-$) parities under $\mathbb{C}_{2x}$.
Hence Eq. (\ref{Coupled_LLG_most_general}) becomes
	\begin{equation}\label{Coupled_LLG_EvenOddParity_general_inplane_dc_field}
		\begin{split}
			\mathrm{i}\Omega\delta\mathbf{m}_{1(2)}^{+}=&\mathbf{m}_{A,1(2)}^{\mathrm{eq}}\times\left\{\left(\delta\mathbf{m}_{1}^{+}+\delta\mathbf{m}_{2}^{+}\right)+k(\delta\mathbf{m}_{1(2)}^{+}\cdot\mathbf{e}_z)\mathbf{e}_z  \right.  \\ 
			&\qquad\qquad +p\left(\mathbf{m}_A^{\mathrm{eq}}\cdot\mathbf{m}_B^{\mathrm{eq}}\right)\mathbb{C}_{2x}\left(\delta\mathbf{m}_{1}^{+}+\delta\mathbf{m}_{2}^{+}\right)  \\
			&\qquad\qquad
			\left.+2p\left[\mathbf{m}_B^{\mathrm{eq}}\cdot\left(\delta\mathbf{m}_{1}^{+}+\delta\mathbf{m}_{2}^{+}\right)\right]\mathbf{m}_{B}^{\mathrm{eq}}\right\},   \\
			\mathrm{i}\Omega\delta\mathbf{m}_{1(2)}^{-}=&\mathbf{m}_{A,1(2)}^{\mathrm{eq}}\times\left\{\left(\delta\mathbf{m}_{1}^{-}+\delta\mathbf{m}_{2}^{-}\right)+k(\delta\mathbf{m}_{1(2)}^{-}\cdot\mathbf{e}_z)\mathbf{e}_z  \right.  \\ 
			&\qquad\qquad
			\left.-p\left(\mathbf{m}_A^{\mathrm{eq}}\cdot\mathbf{m}_B^{\mathrm{eq}}\right)\mathbb{C}_{2x}\left(\delta\mathbf{m}_{1}^{-}+\delta\mathbf{m}_{2}^{-}\right)\right\}.
		\end{split}
	\end{equation}

In the local coordinate system
``$(\mathbf{e}_{\mathbf{m}}^{1(2)}\equiv\mathbf{m}_{A,1(2)}^{\mathrm{eq}},\mathbf{e}_{\phi}^{1(2)}\equiv\mathbf{e}_{z}\times\mathbf{e}_{\mathbf{m}}^{1(2)},\mathbf{e}_{z})$" we decompose $\delta\mathbf{m}_{1(2)}^{\pm}$
as $\delta\mathbf{m}_{1(2)}^{\pm}=\delta m_{1(2),\phi}^{\pm}\mathbf{e}_{\phi}^{1(2)}+\delta m_{1(2),z}^{\pm}\mathbf{e}_{z}$.
Then Eq. (\ref{Coupled_LLG_EvenOddParity_general_inplane_dc_field}) is decoupled into two $4\times 4$ 
matrix equations.
After diagonalizing them, the eigenfrequencies and eigenvectors of spin waves with different parities are obtained.

\subsubsection{IV.B.2 Subspace with even parity under $\mathbb{C}_{2x}$}
The  coupled vectorial equations with even parity are transformed into their matrix-form counterpart:
\begin{equation}\label{InPlaneDC_Positive_parity_4x4_general}
	\Omega\left(\begin{array}{c}
		\delta m_{1,\phi}^+   \\
		\delta m_{2,\phi}^+   \\
		\mathrm{i}\delta m_{1,z}^+   \\
		\mathrm{i}\delta m_{2,z}^+   \\
	\end{array}\right)
	=P\left(\begin{array}{c}
		\delta m_{1,\phi}^+   \\
		\delta m_{2,\phi}^+   \\
		\mathrm{i}\delta m_{1,z}^+   \\
		\mathrm{i}\delta m_{2,z}^+   \\
	\end{array}\right).
\end{equation}
The $4\times 4$ matrix $P$ is partitioned and antidiagonal.
By defining $w\equiv p[\cos2\phi_1+\cos2\phi_2+2\cos(\phi_1-\phi_2)]$, the nonzero components of $P$ are:
$P_{1,4}=P_{2,3}=1-w$, $P_{1,3}=P_{2,4}=1-w+k$,
$P_{3,1}=1-w\cos2\phi_1+2p[\sin(\phi_2-\phi_1)-\sin2\phi_1]^2\equiv r$,
$P_{4,2}=1-w\cos2\phi_2+2p[\sin(\phi_2-\phi_1)+\sin2\phi_2]^2\equiv t$,
and $P_{3,2}=P_{4,1}=\cos(\phi_1+\phi_2)-w\cos(\phi_1-\phi_2)+2p[\sin(\phi_2-\phi_1)+\sin2\phi_2][\sin(\phi_2-\phi_1)-\sin2\phi_1]\equiv s$.
Suppose $P$ is diagonalized with the form: $P\cdot S=S\cdot\mathrm{diag}\{\Omega_1,\Omega_2,\Omega_3,\Omega_4\}$, 
where $\Omega_{k}$ and $S_{*,k}$ are the eigenvalue (normalized resonance frequency)  and 
the corresponding eigenvector, respectively.
Therefore, the spin wave vector for $\Omega_k$ is:
$\delta\mathbf{m}_1^+ + \delta\mathbf{m}_2^+ \propto (S_{1,k} \mathbf{e}_{\phi}^1-\mathrm{i}S_{3,k}\mathbf{e}_z)+(S_{2,k}\mathbf{e}_{\phi}^2-\mathrm{i}S_{4,k}\mathbf{e}_z)$.
If $S_{3,k}/S_{1,k}$ is real, $\delta\mathbf{m}_1^+$ is elliptically polarized.
We then define $S_{3,k}/S_{1,k}$ as the corresponding ellipticity with positive (negative)
value indicating right-handed (left-handed) polarity.
The same convention applies to $\delta\mathbf{m}_2^+$.

For SF phase ($\phi_1=\phi_2$), Eq. (\ref{Symmetrical_Flower_theta1_phi1}) gives
$\cos\phi_1=-(\sqrt[3]{Y_+}+\sqrt[3]{Y_-})/(24p)$ with $Y_{\pm}=96p[-9ph\pm\sqrt{3(27p^2h^2+4p)}]$.
Then the secular equation, $\det |P-I_{4\times4}\Omega|=0$, provides:
$\Omega_1=-\Omega_2=\sqrt{k\sin\phi_1\tan\phi_1(3h-4\cos\phi_1)}$ and
$\Omega_3=-\Omega_4=\sqrt{(4\cos\phi_1-h)[(k+4)\cos\phi_1-h]}$.
When $p<1/4$, $\Omega_{1\sim 4}$ always exist for $h\in(0,h^{\mathrm{FM}})$
and the matrix $S$ reads,
\begin{equation}\label{InPlaneDC_Positive_parity_S_sf_phase}
	S=\left[\begin{array}{cccc}
		1&  1& 1& 1 \\
		-1& -1& 1& 1 \\
		u& -u& v& -v \\
		-u&  u& v& -v \\
	\end{array}\right],
\end{equation}
with $u\equiv\Omega_1/k>0$ and $v\equiv\Omega_3/|4+k-h/\cos\phi_1|>0$.
For $\Omega_1$ branch, $\delta \mathbf{m}_{1}^{+}\propto \mathbf{e}_{\phi}^1-\mathrm{i}u\mathbf{e}_z$
and $\delta \mathbf{m}_{2}^{+}\propto -\mathbf{e}_{\phi}^2+\mathrm{i}u\mathbf{e}_z$
are both right-handed polarized, bearing the same ellipticity $u$ when respectively facing to $+\mathbf{e}_{\mathbf{m}}^{1}$
and $+\mathbf{e}_{\mathbf{m}}^{2}$. Meantime, they bear a fixed $\pi$ difference in their phases. 
While for $\Omega_2$ branch, $\delta \mathbf{m}_{1,2}^{+}$ become left-handed polarized meantime
the absolute value of ellipticity and phase difference are unchanged.
As For $\Omega_3$ ($\Omega_4$) branch, both $\delta \mathbf{m}_{1}^{+}$ and $\delta \mathbf{m}_{2}^{+}$ are 
right-handed (left-handed) with the same ellipticity $v$ ($-v$) and zero phase difference. 
For all branches, $|\delta \mathbf{m}_{1}^{+}|^2=|\delta \mathbf{m}_{2}^{+}|^2$ always holds, 
implying that the spin waves in SF phase have the same weight on the two sublattices in each layer.
When $p>1/4$, $\Omega_{1,2}$ still persist for $h\in(0,h^{\mathrm{FM}})$,
but $\Omega_{3,4}$ disappear when $h$ falls into $(h_\mathrm{min},h_c)$
with $h_c\equiv\frac{k+4}{2}\sqrt{\frac{k+2}{2p}}>h_\mathrm{max}$.
Correspondingly, the matrix $S$ becomes,
\begin{equation}\label{InPlaneDC_Positive_parity_S_sf_phase_p_gt_quarter}
	S=\left\{\begin{array}{c}
		\left[\begin{array}{cccc}
			1&  1& 1& 1 \\
			-1& -1& 1& 1 \\
			u& -u& v& -v \\
			-u&  u& v& -v \\
		\end{array}\right],\ \  h\in \left(0,h_{_\mathrm{min}}\right]\qquad\ \\
		\\
		\left[\begin{array}{cccc}
			1&  1& 1& 1 \\
			-1& -1& 1& 1 \\
			u& -u& -v& v \\
			-u&  u& -v& v \\
		\end{array}\right],\ \   h\in\left[h_c, h^{\mathrm{FM}}\right).\ \  \\
	\end{array}
	\right.
\end{equation}
For $h\in\left[h_c, h^{\mathrm{FM}}\right)$, 
the $\Omega_{3,4}$ branches (if exist) exchange the eigenvectors while leaving the rest unchanged.

Then we turn to MRS phase ($|\phi_1|<\phi_2<\pi-|\phi_1|$) which only exists in
$[h_{\mathrm{min}},h_{\mathrm{max}}]$ when $p>1/4$. 
Now $w\equiv1$ thus $P_{1,4}=P_{2,3}=0$, $P_{1,3}=P_{2,4}=k$, $t>r>0$ and $rt>s^2$.
The eigenfrequencies become: 
$\Omega^{\prime}_1=-\Omega^{\prime}_2=\sqrt{(k/2)[(t+r)+\sqrt{(t-r)^2+4s^2}]}$,
and $\Omega^{\prime}_3=-\Omega^{\prime}_4=\sqrt{(k/2)[(t+r)-\sqrt{(t-r)^2+4s^2}]}$,
which always exist for $h\in[h_{\mathrm{min}},h_{\mathrm{max}}]$.
The corresponding eigenvector matrix reads
\begin{equation}\label{InPlaneDC_Positive_parity_S_af_phase}
	S'=\left[\begin{array}{cccc}
		1&  1& 1& 1 \\
		\frac{1}{\rho}& \frac{1}{\rho}& -\rho& -\rho \\
		u'& -u'& v'& -v'\\
		\frac{u'}{\rho} &  -\frac{u'}{\rho}& -\rho v'& \rho v'\\
	\end{array}\right],
\end{equation}
with $u'\equiv \Omega'_1/k$, $v'\equiv\Omega'_3/k$, and $\rho\equiv [\sqrt{(t-r)^2+4s^2}-(t-r)]/(2s)$.
For positive (negative) eigenfrequencies $\Omega'_{1,3}$ ($\Omega'_{2,4}$), the 
corresponding spin waves are right-handed (left-handed).
For $\Omega'_{1,2}$ ($\Omega'_{3,4}$) branches, the ellipticity of spin waves is $\pm u'$ ($\pm v'$),
meantime $|\delta \mathbf{m}_{1}^{+}|^2/|\delta \mathbf{m}_{2}^{+}|^2$ is $\rho^2$ ($1/\rho^2$).
In particular, when $h\rightarrow h_s=\sqrt{\frac{4}{3p}-\frac{2}{3}(\sqrt[3]{Y_+}+\sqrt[3]{Y_-})}$
with $Y_{\pm}=p^{-3}-72p^{-2}\pm 3p^{-2}\sqrt{6(94-4p-3p^{-1})}$, $s\rightarrow 0$.
Then $\rho\approx s/(t-r)\rightarrow 0$, leading to the dominant $\delta\mathbf{m}_2^+$
($\delta\mathbf{m}_1^+$) spin wave components for $\Omega'_{1,2}$ ($\Omega'_{3,4}$) branches.
At last, the phase difference between $\delta \mathbf{m}_{1,2}^{+}$ depends on $h$.
For $h_{\mathrm{min}}<h<h_s$, $\rho<0$ hence the phase difference is $\pi$ for $\Omega'_{1,2}$ 
and $0$ for $\Omega'_{3,4}$, respectively.
While for $h_s<h<h_{\mathrm{max}}$, $\rho$ becomes positive and
the phase difference changes to 0 ($\pi$) for $\Omega'_{1,2}$ ($\Omega'_{3,4}$).

\subsubsection{IV.B.3 Subspace with odd parity under $\mathbb{C}_{2x}$}
The matrix-form spin-wave equation for odd parity reads
\begin{equation}\label{InPlaneDC_Negative_parity_4x4_general}
	\Omega\left(\begin{array}{c}
		\delta m_{1,\phi}^-   \\
		\delta m_{2,\phi}^-   \\
		\mathrm{i}\delta m_{1,z}^-   \\
		\mathrm{i}\delta m_{2,z}^-   \\
	\end{array}\right)
	=Q\left(\begin{array}{c}
		\delta m_{1,\phi}^-   \\
		\delta m_{2,\phi}^-   \\
		\mathrm{i}\delta m_{1,z}^-   \\
		\mathrm{i}\delta m_{2,z}^-   \\
	\end{array}\right).
\end{equation}
The $4\times 4$ matrix $Q$ is also completely antidiagonal, with the nonzero components
$Q_{1,3}=Q_{2,4}=1+k+w$, $Q_{1,4}=Q_{2,3}=1+w$, $Q_{3,1}=1+w\cos2\phi_1\equiv \tilde{r}$, 
$Q_{4,2}=1+w\cos2\phi_2\equiv \tilde{t}$,
and $Q_{3,2}=Q_{4,1}=\cos(\phi_1+\phi_2)+w\cos(\phi_1-\phi_2)\equiv \tilde{s}$.
It can be further diagonalized as: $Q\cdot \tilde{S}=\tilde{S}\cdot\mathrm{diag}\{\tilde{\Omega}_1,\tilde{\Omega}_2,\tilde{\Omega}_3,\tilde{\Omega}_4\}$.

For SF phase, $\tilde{\Omega}_1=-\tilde{\Omega}_2=\sqrt{k\sin\phi_1\tan\phi_1(4\cos\phi_1-h)}$,
$\tilde{\Omega}_3=-\tilde{\Omega}_4=\sqrt{h(k\cos\phi_1+h)}$.
When $p<1/4$, they always exist when $h\in(0,h^{\mathrm{FM}})$.
Then $\tilde{S}$ reads
\begin{equation}\label{InPlaneDC_Negative_parity_S_sf_phase}
	\tilde{S}=\left[\begin{array}{cccc}
		1&  1& 1& 1 \\
		-1& -1& 1& 1 \\
		\tilde{u}& -\tilde{u}& \tilde{v}& -\tilde{v} \\
		-\tilde{u}&  \tilde{u}& \tilde{v}& -\tilde{v} \\
	\end{array}\right],
\end{equation}
with $\tilde{u}\equiv\tilde{\Omega}_1/k>0$ and $\tilde{v}\equiv\tilde{\Omega}_3/(k+h/\cos\phi_1)>0$.
Again for all positive (negative) eigenfrequencies $\tilde{\Omega}_{1,3}$ ($\tilde{\Omega}_{2,4}$), 
the corresponding spin waves are right-handed (left-handed).
Meantime, the phase difference between $\delta \mathbf{m}_{1,2}^{-}$
for $\tilde{\Omega}_{1,2}$ ($\tilde{\Omega}_{3,4}$) is $\pi$ ($0$).
When $p$ is larger than $1/4$, $\tilde{\Omega}_{1,2}$ (and the corresponding eigenvectors) only exist
for $h\in(0,h_\mathrm{min}]$, while $\tilde{\Omega}_{3,4}$ persist.

For MRS phase, $w\equiv1$ thus $Q_{1,4}=Q_{2,3}=2$ and $Q_{1,3}=Q_{2,4}=2+k$.
Now $\tilde{r}>\tilde{t}>0$ and $\tilde{r}\tilde{t}=\tilde{s}^2$.
In particular, $\tilde{s}>0$ for $p<2$ which is reasonable for real all-antiferromagnetic junctions.
The eigenfrequencies then become: 
$\tilde{\Omega}^{\prime}_1=\tilde{\Omega}^{\prime}_2=0$,
and $\tilde{\Omega}^{\prime}_3=-\tilde{\Omega}^{\prime}_4=\sqrt{(k+2)(\tilde{r}+\tilde{t})+4\tilde{s}}$,
which always exist for $h\in[h_{\mathrm{min}},h_{\mathrm{max}}]$.
Correspondingly, the eigenvector matrix is
\begin{equation}\label{InPlaneDC_Positive_parity_S_af_phase}
	\tilde{S}'=\left[\begin{array}{cccc}
		1&  1& 1& 1 \\
		-\tilde{\rho}& -\tilde{\rho}& \tilde{\lambda}& \tilde{\lambda} \\
		0& 0& \tilde{u}' & -\tilde{u}' \\
		0& 0& \tilde{\lambda}\tilde{v}' & -\tilde{\lambda}\tilde{v}' \\
	\end{array}\right],
\end{equation}
with $\tilde{\rho}\equiv [(2+k)\tilde{r}+2\tilde{s}]/[(2+k)\tilde{s}+2\tilde{t}]$, 
$\tilde{\lambda}\equiv [(2+k)\tilde{t}+2\tilde{s}]/[(2+k)\tilde{s}+2\tilde{t}]$,
$\tilde{u}'=\tilde{\Omega}'_3\tilde{s}/[(2+k)\tilde{s}+2\tilde{t}]$,
and $\tilde{v}'=\tilde{\Omega}'_3\tilde{t}/[(2+k)\tilde{t}+2\tilde{s}]$.
For $\tilde{\Omega}'_{1,2}=0$, $\delta\mathbf{m}_{1}^-$ ($\delta\mathbf{m}_{2}^-$) 
only has time-independent component along $\mathbf{e}_{\phi}^{1}$ ($\mathbf{e}_{\phi}^{2}$) in $xy-$plane.
On the other hand, for $\tilde{\Omega}'_3$, the spin wave components 
$\delta \mathbf{m}_{1}^-$ and $\delta \mathbf{m}_{2}^-$ are both right-handed polarized
with synchronous phases and different ellipticity ($\tilde{u}'$ and $\tilde{v}'$, respectively).
Similar situation holds for $\tilde{\Omega}'_4$, the only difference is the polarization changes to left hand.
At last, for $\tilde{\Omega}'_{3,4}$ branches, 
$|\delta \mathbf{m}_{1}^{+}|^2/|\delta \mathbf{m}_{2}^{+}|^2=\tilde{u}'/(\tilde{\lambda}^2\tilde{v}')$, 
implying that the spin waves in MRS phase with odd parity have different weights on the two sublattices in each layer.

\subsubsection{IV.B.4 Numerical examples and asymptotics}
\begin{figure} [htbp]
	\centering
	\includegraphics[width=0.43\textwidth]{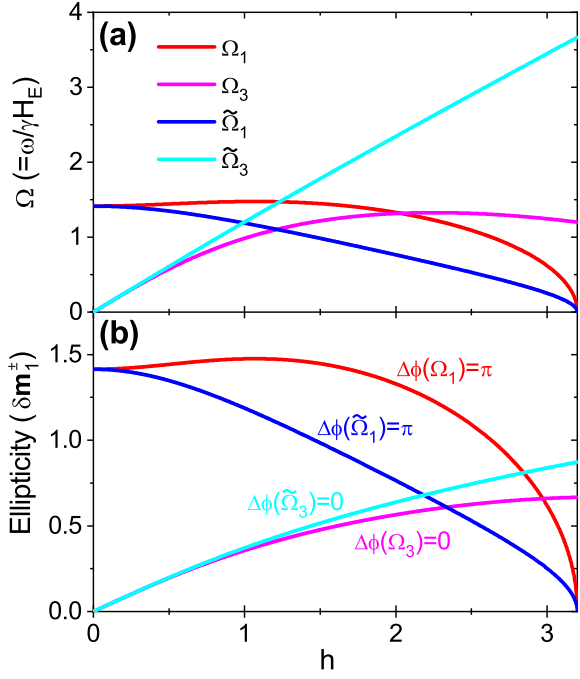}
	\caption{(Color online) Eigenfrequencies, ellipticities, and phase differences of
		spin wave components $\delta\mathbf{m}^{\pm}_{1,2}$ in symmetrical all-antiferromagnetic junctions
		with $k=1$ and $p=0.15$ under in-plane dc fields $h\in \left(0,h^{\mathrm{FM}}\right)$. 
		The equilibrium magnetization layout falls into the SF phase.
		(a) Four positive eigenfrequencies: $\Omega_{1}$ (red), $\Omega_{3}$ (magenta),
		$\tilde{\Omega}_1$ (blue) and $\tilde{\Omega}_3$ (cyan).
		(b) Ellipticities of $\delta\mathbf{m}^{+}_{1}$ for $\Omega_{1,3}$ and
		$\delta\mathbf{m}^{-}_{1}$ for $\tilde{\Omega}_{1,3}$, 
		which are identical to those of $\delta\mathbf{m}^{\pm}_{2}$. The corresponding phase differences 
		between $\delta\mathbf{m}^{\pm}_{1,2}$ for all four branches are indicated. 
	}\label{fig3}
\end{figure}

To be more intuitive, the eigenfrequencies and corresponding spin-wave features 
with $k=1$ and two typical values of $p$ are calculated.
In Fig. \ref{fig3}, $p=0.15$ hence SF phase is the only choice for the ground state. 
The four positive eigenfrequency branches, $\Omega_{1,3}$ with even parity and
$\tilde{\Omega}_{1,3}$ with odd parity, are plotted in Fig. \ref{fig3}(a).
First, $\Omega_1$ (red) and $\tilde{\Omega}_1$ (blue) both start from $\sqrt{2k}$ at $h=0$  
but have different concavity: for $h=\iota\ll 1$, 
\begin{equation}\label{InPlaneDC_SF_phase_Omega1_tildeOmega1_h_0_expansion}
\begin{split}
	\Omega_1\approx& \sqrt{2k}\left[1+\frac{1}{2}\left(p-\frac{1}{4}\right)\iota^2\right],  \\
	\tilde{\Omega}_1\approx& \sqrt{2k}\left[1-\frac{1}{2}\left(p+\frac{1}{4}\right)\iota^2\right].
\end{split}
\end{equation}
Then they converge to 0 at $h=h^{\mathrm{FM}}$ as
\begin{equation}\label{InPlaneDC_SF_phase_Omega1_tildeOmega1_h_FM_expansion}
	\begin{split}
		\Omega_1 \approx& 2\sqrt{k(1+12p)f(p)}\sqrt{\iota},  \\
		\tilde{\Omega}_1\approx& 2\sqrt{k(1-4p)f(p)}\sqrt{\iota},
	\end{split}
\end{equation}
with $\iota\equiv h^{\mathrm{FM}}-h$, and $f(p)\equiv[\sqrt[3]{\sqrt{(h^{\mathrm{FM}})^2+\frac{4}{27p}}-h^{\mathrm{FM}}}+\sqrt[3]{2p}]/[3\sqrt[3]{2p}\sqrt{(h^{\mathrm{FM}})^2+\frac{4}{27p}}]$. 
As for $\Omega_3$ (magenta) and $\tilde{\Omega}_3$ (cyan), around $h=0$ they both linearly increases 
with the same slope $\sqrt{(k+2)/2}$, and have no other zero points.
In particular, $\tilde{\Omega}_3$ is approximately linear with $h$ in the full range of $h\in(0,h^{\mathrm{FM}})$.
As a consequence, we have totally four accidental crossings (two inter-parity and two intra-parity) 
in the resonance spectrum.
In Fig. \ref{fig3}(b), the corresponding ellipticities of $\delta\mathbf{m}^{\pm}_1$ 
for the four branches (``$+$" for $\Omega_{1,3}$, ``$-$" for $\tilde{\Omega}_{1,3}$) from 
Fig. \ref{fig3}(a) are plotted. Note that for $p<1/4$, ellipticities of $\delta\mathbf{m}^{\pm}_2$
coincide with those of $\delta\mathbf{m}^{\pm}_1$ for each branch so that we have not provided.
For $\Omega_1$ and $\tilde{\Omega}_1$, ellipticities of $\delta\mathbf{m}^{\pm}_1$ decreases from
$\sqrt{2/k}$ (at $h=0$) to 0 (at $h^{\mathrm{FM}}$), implying a transition from circular to linear (in $xy-$plane)
polarization of $\delta\mathbf{m}^{\pm}_1$ as $h$ increases.
Accordingly, the phase difference between $\delta\mathbf{m}^{\pm}_1$ and $\delta\mathbf{m}^{\pm}_2$ 
is always $\pi$, as indicated in Fig. \ref{fig3}(b).
Alternatively, for $\Omega_3$ and $\tilde{\Omega}_3$,ellipticities of $\delta\mathbf{m}^{\pm}_1$ 
increases from 0 (at $h=0$) to respectively $\sqrt{(1-4p)/(1-4p+k/2)}$ and $\sqrt{(1+4p)/(1+4p+k/2)}$ 
(at $h^{\mathrm{FM}}$), indicating a linear to circular polarization transition of spin waves. 
Meantime,  the phase difference between $\delta\mathbf{m}_1^{\pm}$ and $\delta\mathbf{m}_2^{\pm}$ changes to 0.
At last, the spin-wave intensity ratio, $|\delta\mathbf{m}_1^{\pm}|^2/|\delta\mathbf{m}_2^{\pm}|^2$, 
always equals to 1 for all frequency branches, implying a equal-weighting of spin-waves in SF phase.

\begin{figure} [htbp]
	\centering
	\includegraphics[width=0.42\textwidth]{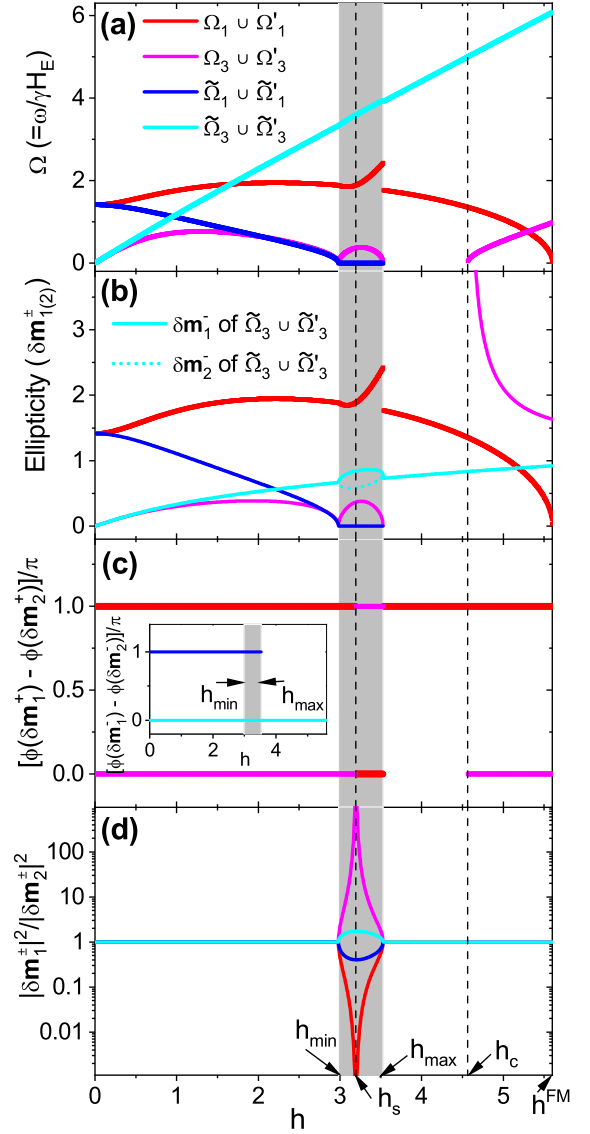}
	\caption{(Color online) Eigenfrequencies, ellipticities, phase differences and weight ratios of
		spin wave components in symmetrical all-antiferromagnetic junctions
		with $k=1$ and $p=0.45$ under in-plane dc fields with $h\in \left(0,h^{\mathrm{FM}}\right)$. 
		The equilibrium magnetization layout can be either SF (white) or MRS (shaded) phase.
		(a) Four positive eigenfrequencies $\Omega_{1}$ (red),  $\Omega_{3}$ (magenta), $\tilde{\Omega}_1$ (blue),
		and $\tilde{\Omega}_3$ (cyan).
		(b) Solid (dashed) curves are the ellipticities of $\delta\mathbf{m}^{\pm}_{1}$ ($\delta\mathbf{m}^{\pm}_{2}$) 
		for all four branches. 
		(c) Phase differences between $\delta\mathbf{m}^{+}_{1,2}$ for $\Omega_{1,3}$ branches. Those
		for  $\tilde{\Omega}_{1,3}$ are provided in the inset.
		(d) Intensity ratio of the spin wave component on group 1 over that on group 2. In SF phase (white area)
		the data from all four branches (if exist) coincide, while in MRS phase (shaded area) they split.
	}\label{fig4}
\end{figure}

In Fig. \ref{fig4}, $p$ increases to $0.45$ hence either SF or MRS phase can be the ground state,
which has been indicated respectively by white or shaded areas.
Based on them, $\Omega_{1}\cup\Omega'_{1}$, $\Omega_{3}\cup\Omega'_{3}$,
$\tilde{\Omega}_{1}\cup\tilde{\Omega}'_{1}$ and $\tilde{\Omega}_{3}\cup\tilde{\Omega}'_{3}$ 
are provided in Fig. \ref{fig4}(a).
Here branches without (with) a prime are in SF (MRS) phase.
Still, two inter-parity (red-cyan, blue-magenta) and two intra-parity (red-magenta, blue-cyan) crossings exist 
(blue-magenta crossing occurs around $h=2.075$ and is not very clear in this dimension).
The asymptotic behaviors of these four branches around $h=0$ are the same as Eq. (\ref{InPlaneDC_SF_phase_Omega1_tildeOmega1_h_0_expansion}), however due to the emergence of 
MRS phase in $[h_{\mathrm{min}},h_{\mathrm{max}}]$ with $h_{\mathrm{min}}=2.9814$ and
$h_{\mathrm{max}}=3.5277$, several interesting changes occurs.
Recalling that at $h=h_{\mathrm{min}}$ ($h_{\mathrm{max}}$) the equilibrium system undergoes a 
second-order (first-order) PT, hence both $\Omega_{1}\cup\Omega'_{1}$ 
and $\tilde{\Omega}_3\cup\tilde{\Omega}'_3$ are continuous 
(discontinuous) at $h=h_{\mathrm{min}}$ ($h_{\mathrm{max}}$). 
Except for this, $\Omega_{1}$ and $\tilde{\Omega}_3$ do not change much compared with 
the counterparts under $p=0.15$.
The other two, however, have changed dramatically.
For $\Omega_3\cup\Omega'_3$ (magenta), three extra zero points ($h_{\min}$, $h_{\max}$ and $h_c=4.5644 $) emerge, with
the following asymptotic behaviors
\begin{equation}\label{InPlaneDC_Omega3_hmin_hmax_hc_expansions}
	\left\{\begin{array}{c}
		\iota=\left\{\begin{array}{c}
			    h_{\mathrm{min}}-h:\ \Omega_3=\sqrt{\frac{1}{4}k h_{\mathrm{min}}}\sqrt{\iota},\qquad\qquad\qquad
				\\
			    h-h_{\mathrm{min}}:\ \Omega'_3=\sqrt{\frac{22p-1}{16p-4}k h_{\mathrm{min}}}\sqrt{\iota},\qquad\qquad
	    	  \end{array}
       	      \right.
		\\
		\iota=h_{\mathrm{max}}-h:\ \Omega'_3=\sqrt{\frac{4p}{8p+3}k h_{\mathrm{max}}}\sqrt{\iota},\qquad\qquad\qquad
		\\
		\iota=h-h_c:\ \Omega_3=2\sqrt{h_c [1+(k+4)g(p)]}\sqrt{\iota},
	\end{array}
	\right.
\end{equation}
where $0<\iota\ll 1$, and $g(p)\equiv[\sqrt[3]{\sqrt{(h_c)^2+\frac{4}{27p}}-h_c}+\frac{1}{2}\sqrt{\frac{k+2}{2p}}\sqrt[3]{2p}]/[3\sqrt[3]{2p}\sqrt{(h_c)^2+\frac{4}{27p}}]$. 
While for $\tilde{\Omega}_1\cup\tilde{\Omega}'_1$ (blue), the part in SF phase shrinks to $(0,h_{\mathrm{min}})$
with the following asymptotic behavior on the left hand of $h_{\mathrm{min}}$,
\begin{equation}\label{InPlaneDC_SF_phase_tildeOmega1_hmin_expansion}
		\tilde{\Omega}_1\approx  \sqrt{k\frac{16-h_{\mathrm{min}}^2}{4h_{\mathrm{min}}}} \sqrt{\iota},\quad \iota=h_{\mathrm{min}}-h.
\end{equation}
In addition, the zero flat $\tilde{\Omega}'_1$ spectrum in MRS phase indicates a time-independent
magnetization deviation in $xy-$plane from the equilibrium layout.
In 2021, Yuan and Duine revealed that the resonance frequency $\omega$ follows a universal power law  
$\omega\propto|H-H_c|^p$, where $H_c$ is the critical field at which the resonance frequency is zero\cite{Yuan_PRB_2021}. 
When the magnet preserves rotational symmetry around the external field $\mathbf{H}$, $p=1$, otherwise, $p=1/2$.
All of our asymptotic results on eigenfrequencies [Eqs. (\ref{InPlaneDC_SF_phase_Omega1_tildeOmega1_h_FM_expansion}) 
to (\ref{InPlaneDC_SF_phase_tildeOmega1_hmin_expansion})] confirm their claims. 
In addition, they asserted that the zero frequency is often accompanied by a reorientation transition in the magnetization. 
For the case in which $p=1/2$, this transition is described by a Landau theory for second-order PTs.
In our symmetric all-antiferromagnetic junctions, the PT at $h=h_{\mathrm{min}}$ (SF$\leftrightarrow$MRS)
is applicable to this conclusion, however the one at $h=h_{\mathrm{max}}$ (MRS$\leftrightarrow$SF) breaks this rule.
At last, inter-parity and intra-parity crossings still coexist.

In Fig. \ref{fig4}(b), ellipticities of $\delta\mathbf{m}^{\pm}_{1,2}$ are provided.
For the entire region of $h\in(0,h^{\mathrm{FM}})$, 
ellipticities of $\delta\mathbf{m}_2$ coincide with those of $\delta\mathbf{m}_1$
for $\Omega_{1}\cup\Omega'_{1}$, $\Omega_{3}\cup\Omega'_{3}$, and $\tilde{\Omega}_{1}\cup\tilde{\Omega}'_{1}$
so that we only provide the latter.
As for $\tilde{\Omega}_{3}\cup\tilde{\Omega}'_{3}$, inconsistency occurs in MRS phase which has been
indicated by the departure of solid and dotted cyan curves.
On one hand, the two branches with odd parity ($\tilde{\Omega}_{1}\cup\tilde{\Omega}'_{1}$
and $\tilde{\Omega}_{3}\cup\tilde{\Omega}'_{3}$) are relatively simple.
Ellipticity of $\delta\mathbf{m}_{1,2}^{-}$ from the $\tilde{\Omega}_{1}\cup\tilde{\Omega}'_{1}$
($\tilde{\Omega}_{3}\cup\tilde{\Omega}'_{3}$) branch decreases (increases) from $\sqrt{2/k}$ (0)
at $h=0$ to 0 ($\sqrt{(1+4p)/(1+4p+k/2)}$) at $h=h_{\mathrm{min}}$ ($h_{\mathrm{max}}$), 
indicating a gradual evolution from circular (linear) to linear (circular) polarization of 
$\delta\mathbf{m}_{1,2}^{-}$ as $h$ increases.
Meantime, the phase difference between $\delta\mathbf{m}_{1,2}^{-}$ is fixed as $\pi$ (0) whenever 
the eigenfrequency exists, as shown in the inset of Fig. \ref{fig4}(c).
The intensity ratio, $|\delta\mathbf{m}_{1}^-|^2/|\delta\mathbf{m}_{2}^-|^2$, keeps 1 in SF phase for
both $\tilde{\Omega}_{1}\cup\tilde{\Omega}'_{1}$ and $\tilde{\Omega}_{3}\cup\tilde{\Omega}'_{3}$ branches,
and separates in MRS phase [see Fig. \ref{fig4}(d)].
Clearly, in the $\tilde{\Omega}'_{1}$ ($\tilde{\Omega}'_{3}$) branch, 
$\delta\mathbf{m}_{2}^-$ ($\delta\mathbf{m}_{1}^-$) dominates.
On the other hand, the two branches with even parity ($\Omega_{1}\cup\Omega'_{1}$ and $\Omega_{3}\cup\Omega'_{3}$) 
are more complex.
The emergence of MRS phase separates the ellipticity of $\Omega_{3}\cup\Omega'_{3}$ branch into four parts and
the corresponding three dividing points are $h_{\mathrm{min}}$, $h_{\mathrm{max}}$ and $h_c$,
as shown in Fig. \ref{fig4}(b).
At the former two (the last one), $\delta\mathbf{m}_{1,2}^+$ become linearly polarized in $xy-$plane ($z-$axis).
As for $\Omega_{1}\cup\Omega'_{1}$ branch, the ellipticity become concave in MRS phase rather than
convex in SF phase, leading to a discontinuity at $h_{\mathrm{max}}$.
Greater interest resides in the intensity ratio in MRS phase.
For $\Omega_{1}\cup\Omega'_{1}$ ($\Omega_{3}\cup\Omega'_{3}$)  branch, spin waves
are concentrated on $\delta\mathbf{m}_2^+$ ($\delta\mathbf{m}_1^+$), as depicted in Fig. \ref{fig4}(d).
In particular, at $h_s=3.1955$ only $\delta\mathbf{m}_2^+$ ($\delta\mathbf{m}_1^+$) exists.
When $h$ exceeds $h_s$, $\delta\mathbf{m}_1^+$ ($\delta\mathbf{m}_2^+$) components reappear, however
the phase difference between $\delta\mathbf{m}_{1,2}^+$ exchanges ($0\leftrightarrow\pi$) 
compared with the case where $h<h_s$ [see Fig. \ref{fig4}(c)].
This reminds us the ``level reversal" behavior in topological insulators after the bulk gap closes and then reopens.

\subsection{IV.C Oblique-dc-field case}
Under in-plane dc fields, the absence of entanglement between subspaces with opposite parity under $\mathbb{C}_{2x}$
leads to the complete decoupling in the dynamic equations of $\delta\mathbf{m}_{1,2}^{+}$ and $\delta\mathbf{m}_{1,2}^{-}$
[see Eq. (\ref{Coupled_LLG_EvenOddParity_general_inplane_dc_field})].
This further results in the accidentally crossing between frequency branches with opposite parity 
in resonance spectrums, implying the absence of magnon-magnon couplings in this system.
To induce it, a reasonable attempt is exerting an oblique dc field with $0<\psi<\pi/2$.
Systematical investigations on the impact of oblique dc fields on coherent magnonics
constitute the main content of this subsection.

\subsubsection{IV.C.1 Emergence of couplings}
For SF phase, $\mathbf{e}_{x^{\prime}}\equiv\mathbf{e}_{x^{\prime\prime}}\ne\mathbf{e}_x$, 
$\mathbf{e}_{y^{\prime}}\equiv\mathbf{e}_{y^{\prime\prime}}=\mathbf{e}_y$ ,
and $\mathbf{e}_{z^{\prime}}\equiv\mathbf{e}_{z^{\prime\prime}}\ne\mathbf{e}_z$,
thus $\mathbb{C}_{2x^{\prime}}=\mathbb{C}_{2x^{\prime\prime}}\equiv \mathbb{C}\ne\mathbb{C}_{2x}$.
In the skewed local coordinate systems
``$(\mathbf{e}_{\mathbf{m}}^{1(2)}\equiv\mathbf{m}_{A,1(2)}^{\mathrm{eq}},\mathbf{e}_{\phi}^{1(2)}\equiv\mathbf{e}_{z'}\times\mathbf{e}_{\mathbf{m}}^{1(2)},\mathbf{e}_{z'})$",
similarly we decompose $\delta\mathbf{m}_{1(2)}^{\pm}$
as $\delta\mathbf{m}_{1(2)}^{\pm}=\delta m_{1(2),\phi}^{\pm}\mathbf{e}_{\phi}^{1(2)}+\delta m_{1(2),z'}^{\pm}\mathbf{e}_{z'}$,
which are the spin wave components with even ($+$) and odd ($-$) parities under $\mathbb{C}$.
Hence Eq. (\ref{Coupled_LLG_most_general}) becomes

	\begin{equation}\label{Coupled_LLG_EvenOddParity_general_outofplane_dc_field_SF_phase}
		\begin{split}
			\mathrm{i}\Omega\delta\mathbf{m}_{1(2)}^{+}=
			&\mathbf{m}_{A,1(2)}^{\mathrm{eq}}\times\left\{\left(\delta\mathbf{m}_{1}^{+}+\delta\mathbf{m}_{2}^{+}\right)\right.  \\ 
			&\qquad
			+2p\left[\mathbf{m}_B^{\mathrm{eq}}\cdot\left(\delta\mathbf{m}_{1}^{+}+\delta\mathbf{m}_{2}^{+}\right)\right]\mathbf{m}_{B}^{\mathrm{eq}}   \\
			&\qquad
			+p\left(\mathbf{m}_A^{\mathrm{eq}}\cdot\mathbf{m}_B^{\mathrm{eq}}\right)\mathbb{C}\left(\delta\mathbf{m}_{1}^{+}+\delta\mathbf{m}_{2}^{+}\right)   \\ 
			&\qquad
			  +\frac{k}{2}\left[\mathbf{e}_z\left(\mathbf{e}_z\cdot\right)+\mathbb{C}\mathbf{e}_z\left(\mathbb{C}\mathbf{e}_z\cdot\right)\right]\delta\mathbf{m}_{1(2)}^{+}  \\
		    &\qquad
			  \left.+\frac{k}{2}\left[\mathbf{e}_z\left(\mathbf{e}_z\cdot\right)-\mathbb{C}\mathbf{e}_z\left(\mathbb{C}\mathbf{e}_z\cdot\right)\right]\delta\mathbf{m}_{1(2)}^{-} 	\right\},   \\
			\mathrm{i}\Omega\delta\mathbf{m}_{1(2)}^{-}=
			&\mathbf{m}_{A,1(2)}^{\mathrm{eq}}\times\left\{\left(\delta\mathbf{m}_{1}^{-}+\delta\mathbf{m}_{2}^{-}\right) \right.  \\ 
			&\qquad
			-p\left(\mathbf{m}_A^{\mathrm{eq}}\cdot\mathbf{m}_B^{\mathrm{eq}}\right)\mathbb{C}\left(\delta\mathbf{m}_{1}^{-}+\delta\mathbf{m}_{2}^{-}\right)    \\
			&\qquad
			  +\frac{k}{2}\left[\mathbf{e}_z\left(\mathbf{e}_z\cdot\right)+\mathbb{C}\mathbf{e}_z\left(\mathbb{C}\mathbf{e}_z\cdot\right)\right]\delta\mathbf{m}_{1(2)}^{-}   \\
		    &\qquad
	    	\left.+\frac{k}{2}\left[\mathbf{e}_z\left(\mathbf{e}_z\cdot\right)-\mathbb{C}\mathbf{e}_z\left(\mathbb{C}\mathbf{e}_z\cdot\right)\right]\delta\mathbf{m}_{1(2)}^{+} \right\}.
		\end{split}
	\end{equation}
Clearly,  $\mathbb{C}\mathbf{e}_z\ne -\mathbf{e}_z$ (since $\mathbf{e}_{x'}\ne\mathbf{e}_{x}$), leading to
the couplings between subspaces with opposite parity under $\mathbb{C}$ and further the possible anticrossings 
in resonance spectrum.

To see more clearly, by defining $\Lambda\equiv(\delta m_{1,\phi}^+,\delta m_{2,\phi}^+,\mathrm{i}\delta m_{1,z'}^+,\mathrm{i}\delta m_{2,z'}^+,\delta m_{1,\phi}^-,\delta m_{2,\phi}^-,\mathrm{i}\delta m_{1,z'}^-,\mathrm{i}\delta m_{2,z'}^-)^{\mathrm{T}}$
the above vectorial equation set becomes
\begin{equation}\label{Coupled_LLG_EvenOddParity_general_outofplane_dc_field_SF_phase_matrix_form}
	 P_{\mathrm{sym}}\Lambda=\Omega \Lambda, 
	\quad P_{\mathrm{asym}}=\left[
	\begin{array}{cc}
		\varTheta  &  \Gamma \\
		\Gamma & \varPhi  \\
	\end{array}\right],
\end{equation}
with
\begin{equation}\label{Coupled_LLG_EvenOddParity_general_outofplane_dc_field_SF_phase_matrice}
\begin{split}
	&\varTheta=\left[\begin{array}{cccc}
		0& 0& a_1+k\cos^2\chi & a_1 \\
		0& 0& a_1 & a_1+k\cos^2\chi \\
		b_1& b_2& 0 & 0 \\
		b_2& b_1& 0 & 0 \\
	\end{array}\right], \\
	&\varPhi=\left[\begin{array}{cccc}
	0& 0& a_2+k\cos^2\chi & a_2 \\
	0& 0& a_2 & a_2+k\cos^2\chi \\
	c_1& c_2& 0 & 0 \\
	c_2& c_1& 0 & 0 \\
   \end{array}\right],  \\
   & \Gamma=\mathrm{i}\times\mathrm{diag}\{-d,d,d,-d\},
\end{split}
\end{equation}
where the definitions of $a_{1,2}$, $b_{1,2}$, $c_{1,2}$ and $d$ are provided in Eq. (\ref{a1_to_f2_definition}).
The nonzero $\psi$ results in finite $d$, hence the nonvanishing coupling matrix $\Gamma$.
In principle, by numerically diagonalizing $P_{\mathrm{sym}}$, the eigenfrequecies, the hybridization of even 
and odd subspaces, the ellipticities and phase difference between spin wave components can be obtained.
Specifically, the secular equation for resonance frequency can be reduced to a standard quartic 
algebraic equation of $\Omega^2$:
$(\Omega^2)^4 + J_3(\Omega^2)^3 +  J_2(\Omega^2)^2 +  J_1(\Omega^2) + J_0 =0$.
The explicit expressions of $J_i$ (functions of $\chi$ and $\epsilon$) have been provided in Appendix A
due to their lengthy forms.
By solving it, the eigenfrequencies in SF phase can be obtained and cross-verified with data
from direct diagonalization.
In addition, the pure-imaginary coupling matrix $\Gamma$ (as long as $\psi>0$)
has two important consequences on the eigenvectors: 
(i)  Nonvanishing hybridization between subspaces with even and odd parities under $\mathbb{C}$ persists.
(ii) There is always a phase difference of $\pm\frac{\pi}{2}$ between components in subspaces with even and odd parities.

For MRS phase, $\mathbf{e}_{x'}\ne\mathbf{e}_{x^{\prime\prime}}$ 
then $\mathbb{C}_{2x'} \mathbf{m}_{A,2}^{\mathrm{eq}}\ne \mathbf{m}_{B,2}^{\mathrm{eq}}$ 
and $\mathbb{C}_{2x^{\prime\prime}} \mathbf{m}_{A,1}^{\mathrm{eq}}\ne \mathbf{m}_{B,1}^{\mathrm{eq}}$. 
The only possible parity operator is the mirror reflection $\mathbb{M}_{xz}$.
We then construct 
$\mathbf{l}_{1(2)}^{\pm}\equiv\delta\mathbf{m}_{A,1(2)} \pm \mathbb{M}_{xz}\delta\mathbf{m}_{B,1(2)}$ 
as the spin wave components with even ($+$) and odd ($-$) parities under $\mathbb{M}_{xz}$.
Note that  
$\mathbb{M}_{xz}(\mathbf{U}\cdot\mathbf{V})=\mathbb{M}_{xz}\mathbf{U}\cdot \mathbb{M}_{xz}\mathbf{V}$
and $\mathbb{M}_{xz}(\mathbf{U}\times\mathbf{V})=-\mathbb{M}_{xz}\mathbf{U}\times \mathbb{M}_{xz}\mathbf{V}$
where $\mathbf{U}$ and $\mathbf{V}$ are arbitrary vectors,
therefore $\mathbb{M}_{xz}$ does not preserve the polarization of spin waves, which is
quite different from the rotation operators $\mathbb{C}_{2x(x',x^{\prime\prime})}$.
Now Eq. (\ref{Coupled_LLG_most_general}) becomes
	\begin{equation}\label{Coupled_LLG_EvenOddParity_general_outofplane_dc_field_af_phase}
	\begin{split}
	\mathrm{i}\Omega\mathbf{l}_{1(2)}^{+}=&\mathbf{m}_{A,1(2)}^{\mathrm{eq}}\times\left\{\left(\mathbf{l}_{1}^{-}+\mathbf{l}_{2}^{-}\right)+k(\mathbf{l}_{1(2)}^{-}\cdot\mathbf{e}_z)\mathbf{e}_z  \right.  \\ 
	&\qquad\qquad \left. -p\left(\mathbf{m}_A^{\mathrm{eq}}\cdot\mathbf{m}_B^{\mathrm{eq}}\right)\mathbb{M}_{xz}\left(\mathbf{l}_{1}^{-}+\mathbf{l}_{2}^{-}\right)\right\},   \\
	\mathrm{i}\Omega\mathbf{l}_{1(2)}^{-}=&\mathbf{m}_{A,1(2)}^{\mathrm{eq}}\times\left\{\left(\mathbf{l}_{1}^{+}+\mathbf{l}_{2}^{+}\right)+k(\mathbf{l}_{1(2)}^{+}\cdot\mathbf{e}_z)\mathbf{e}_z\right.  \\ 
	&\qquad\qquad 
	+p\left(\mathbf{m}_A^{\mathrm{eq}}\cdot\mathbf{m}_B^{\mathrm{eq}}\right)\mathbb{M}_{xz}\left(\mathbf{l}_{1}^{+}+\mathbf{l}_{2}^{+}\right)  \\
	&\qquad\qquad \left.
	+2p\left[\mathbf{m}_B^{\mathrm{eq}}\cdot\left(\mathbf{l}_{1}^{+}+\mathbf{l}_{2}^{+}\right)\right]\mathbf{m}_{B}^{\mathrm{eq}}\right\}.
	\end{split}
	\end{equation}

Recalling that the angle between $\mathbf{e}_{x^{\prime}}$ ($\mathbf{e}_{x^{\prime\prime}}$) 
and $\mathbf{e}_x$ is $\chi$ ($\xi$), meantime the angle between $\mathbf{e}_{x^{\prime}}$ and
$\mathbf{m}_{A,1}^{\mathrm{eq}}$ 
($\mathbf{e}_{x^{\prime\prime}}$ and $\mathbf{m}_{A,2}^{\mathrm{eq}}$) is $\epsilon$ ($\eta$).
For MRS phase under oblique dc fields, generally $0<\chi<\xi<\frac{\pi}{2}-\psi$.
We then denote the plane expanded by $\mathbf{m}_{A,1}^{\mathrm{eq}}$ and $\mathbf{m}_{B,1}^{\mathrm{eq}}$
($\mathbf{m}_{A,2}^{\mathrm{eq}}$ and $\mathbf{m}_{B,2}^{\mathrm{eq}}$) as ``$\chi-$plane ($\xi-$plane)".
On each oblique plane, we define the following local coordinate system: 
(i) $\chi$-plane-based: $(\mathbf{e}_{\mathbf{m}}^{1}\equiv\mathbf{m}_{A,1}^{\mathrm{eq}},\mathbf{e}_{\phi}^{1}\equiv\mathbf{e}_{z^{\prime}}\times\mathbf{e}_{\mathbf{m}}^{1},\mathbf{e}_{z^{\prime}}=-\sin\chi\mathbf{e}_x+\cos\chi\mathbf{e}_z)$, and
(ii) $\xi$-plane-based: $(\mathbf{e}_{\mathbf{m}}^{2}\equiv\mathbf{m}_{A,2}^{\mathrm{eq}},\mathbf{e}_{\phi}^{2}\equiv\mathbf{e}_{z^{\prime\prime}}\times\mathbf{e}_{\mathbf{m}}^{1},\mathbf{e}_{z^{\prime\prime}}=-\sin\xi\mathbf{e}_x+\cos\xi\mathbf{e}_z)$.
Then we decompose $\mathbf{l}_{1(2)}^{\pm}$
as: $\mathbf{l}_{1}^{\pm}=l_{1,\phi}^{\pm}\mathbf{e}_{\phi}^{1}+l_{1,z'}^{\pm}\mathbf{e}_{z^{\prime}}$
and $\mathbf{l}_{2}^{\pm}=l_{2,\phi}^{\pm}\mathbf{e}_{\phi}^{2}+l_{2,z^{\prime\prime}}^{\pm}\mathbf{e}_{z^{\prime\prime}}$.
After defining
$\tilde{\varLambda}\equiv(l_{1,\phi}^+,l_{2,\phi}^+,\mathrm{i}l_{1,z'}^+,\mathrm{i}l_{2,z^{\prime\prime}}^+,l_{1,\phi}^-,l_{2,\phi}^-,\mathrm{i}l_{1,z'}^-,\mathrm{i}l_{2,z^{\prime\prime}}^-)^{\mathrm{T}}$,  
Eq. (\ref{Coupled_LLG_EvenOddParity_general_outofplane_dc_field_af_phase}) is rewritten as
\begin{equation}\label{OutOfPlaneDC_8x8_general}
	P_{\mathrm{asym}}\tilde{\varLambda}=\Omega\tilde{\varLambda},\quad  
	P_{\mathrm{asym}}=\left[
	\begin{array}{cc}
		0_{4\times4} &  \tilde{\varTheta} \\
		\tilde{\varPhi} & 0_{4\times4}  \\
	\end{array}\right],
\end{equation}
in which $\tilde{\varTheta}$ and $\tilde{\varPhi}$ are both $4\times4$ matrices,
with detailed expressions of each element provided in Appendix B.
Clearly, entanglements between subspaces with opposite parity under $\mathbb{M}_{xz}$ emerge, thus 
making the crossings (if exist) in resonance spectrum become anticrossings.
Again, by directly diagonalizing $P_{\mathrm{asym}}$, the eigenfrequecies, eigenvectors and corresponding
features can be obtained.
In particular, the secular equation, $|P_{\mathrm{asym}}-\Omega I_{8\times8}|=0$, is
reduced to $|\tilde{\varTheta}\tilde{\varPhi}-\Omega^2 I_{4\times4}|=0$, which is also 
a quartic algebraic equation of $\Omega^2$.
However the detailed expression is too complicated to be explicitly written out.
One can numerically solve it to cross-verify the data from direct diagonalization.

\subsubsection{IV.C.2 Numerical examples}
To have a direct impression on the oblique-field-induced magnon-magnon coupling, we take 
$\psi=\pi/10$ as an example meantime leaving other parameters ($p$ and $k$) the same as in
Figs. \ref{fig3} and \ref{fig4} (in-plane-dc-field case).
Considering the possible anticrossings in resonance spectrum (thus hybridize the original branches with even and odd 
parities under rotation or mirror reflection operators), in this section we name
the four positive eigenfrequency branches as ``$\Omega_{\mathrm{I}}$, $\Omega_{\mathrm{II}}$, 
$\Omega_{\mathrm{III}}$, and $\Omega_{\mathrm{IV}}$" in the order from highest to lowest around $h=0$.

\begin{figure} [htbp]
	\centering
	\includegraphics[width=0.44\textwidth]{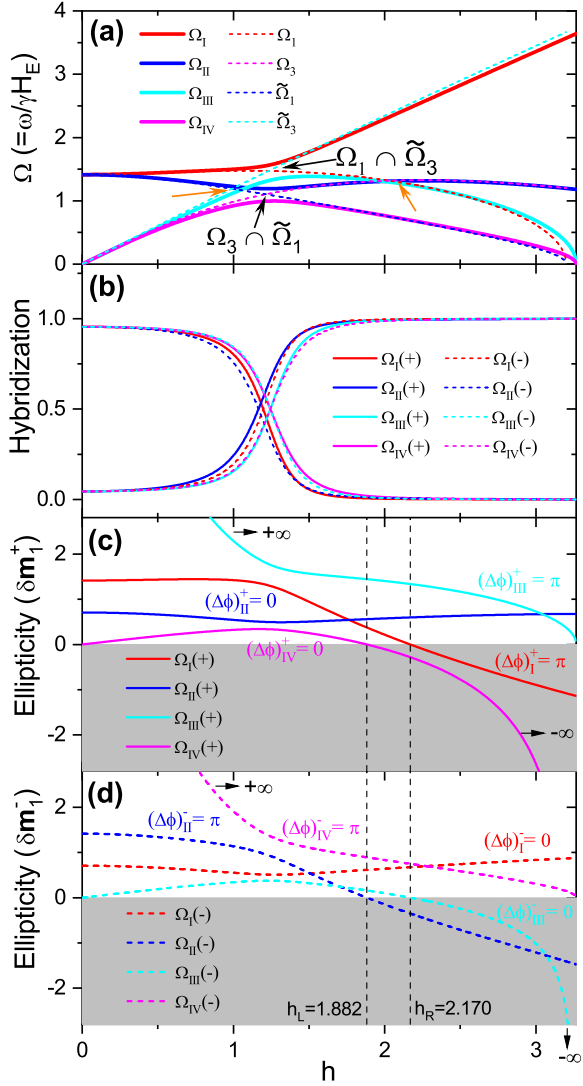}
	\caption{(Color online) Eigenfrequencies, hybridizations, ellipticities, and phase differences of
		spin wave components $\delta\mathbf{m}^{\pm}_{1,2}$ in symmetrical all-antiferromagnetic junctions
		with $k=1$ and $p=0.15$ under oblique dc fields with $\psi=\pi/10$ and $h\in \left(0,h^{\mathrm{FM}}\right)$. 
		The equilibrium magnetization layout falls into the SF phase.
		(a) Four positive eigenfrequencies $\Omega_{\mathrm{I}}$ (red), $\Omega_{\mathrm{II}}$ (blue), 
		$\Omega_{\mathrm{III}}$ (cyan),
		and $\Omega_{\mathrm{IV}}$ (magenta). The dashed curves are $\Omega_{1}$ (red),  $\Omega_{3}$ (magenta), 
		$\tilde{\Omega}_1$ (blue), and $\tilde{\Omega}_3$ (cyan) for in-plane-dc-field cases [see Fig. \ref{fig3}(a)].
		Black arrows indicate the inter-parity crossing points: $\Omega_{1}\cap \tilde{\Omega}_3$ and
		$\tilde{\Omega}_1\cap\Omega_{3}$, while orange arrows indicate the intra-parity ones: 
		$\Omega_{1}\cap \Omega_{3}$ and $\tilde{\Omega}_1\cap\tilde{\Omega}_3$.
		(b) Evolution of hybridization between subspaces with even and odd parities for all four eigenfrequencies.
		Solid (dashed) lines indicate $|\delta\mathbf{m}_1^{+}|^2+|\delta\mathbf{m}_2^{+}|^2$ 
		($|\delta\mathbf{m}_1^{-}|^2+|\delta\mathbf{m}_2^{-}|^2$).
		(c) Ellipticities of $\delta\mathbf{m}^{+}_{1}$ for all four eigenfrequencies, which
		are identical to those of $\delta\mathbf{m}^{+}_{2}$. The corresponding phase differences 
		between $\delta\mathbf{m}^{+}_{1,2}$ are indicated. 
		(d) Counterparts of (c) for $\delta\mathbf{m}^{-}_{1,2}$.
	}\label{fig5}
\end{figure}

For $p=0.15$, only frequency branches in SF phase exist.
In Fig. \ref{fig5}(a) , the four positive eigenfrequencies ($\Omega_{\mathrm{I}\sim\mathrm{IV}}$)
are plotted by solid curves.
Meantime, we repaint the four frequency branches in Fig. \ref{fig3}(a) by dashed curves.
The inter-parity crossing points under in-plane dc fields, $\Omega_{1}\cap \tilde{\Omega}_3$ and
$\tilde{\Omega}_1\cap\Omega_{3}$, are lifted by oblique dc fields and become anticrossings [see black arrows in Fig. \ref{fig5}(a)], 
while the intra-parity ones ($\Omega_{1}\cap \Omega_{3}$ and $\tilde{\Omega}_1\cap\tilde{\Omega}_3$)
persist [orange arrows in Fig. \ref{fig5}(a)].
For each eigenfrequency, its corresponding eigenvector is the superposition of $\delta\mathbf{m}_{1,2}^{\pm}$.
The weights in subspaces with even and odd parities are described by 
$|\delta\mathbf{m}_1^{+}|^2+|\delta\mathbf{m}_2^{+}|^2$ and $|\delta\mathbf{m}_1^{-}|^2+|\delta\mathbf{m}_2^{-}|^2$,
respectively.
The variation of these weights as $h$ increases depicts the evolution of subspace hybridization,
as shown in Fig. \ref{fig5}(b).
Originally, the hybridization is relatively weak when $h\sim 0$ and each eigenvector preserve its
main weight similar to those from in-plane-dc-field case.
When $h$ increases,  around each anticrossing the eigenvector experiences a strong hybridization 
and is then almost evenly distributed in the two subspaces.
As $h$ is further strengthened, the eigenvector transfer its most weight to the subspace with opposite parity.
In addition, the pure-imaginary coupling matrix $\Gamma$ always results in a $\pi/2$ phase difference between 
$\delta\mathbf{m}_1^{+}+\delta\mathbf{m}_2^{+}$ and $\delta\mathbf{m}_1^{-}+\delta\mathbf{m}_2^{-}$.

In Figs. \ref{fig5}(c) and \ref{fig5}(d), the ellipticities of $\delta\mathbf{m}_1^{+}$ and $\delta\mathbf{m}_1^{-}$
from each eigenfrequency branch are plotted, respectively.
In each curve, positive (negative) ellipticity means right-handed (left-handed) elliptical polarization,
and falls into the white (shaded) area.
For magnonics in SF phase, ellipticity of $\delta\mathbf{m}_2^{\pm}$ equals to that of $\delta\mathbf{m}_1^{\pm}$ 
thus we have not provided.
We first focus on the hybridized pair: $\Omega_{\mathrm{I}}$ and $\Omega_{\mathrm{III}}$.
$\delta\mathbf{m}_1^{+}$ of $\Omega_{\mathrm{I}}$ [red solid curve in Figs. \ref{fig5}(c)] 
changes from right-handed around $h=0$ to linear polarized in $x'y$-plane
at $h_{\mathrm{R}}\approx 2.170$ and then to left-handed until $h\rightarrow h^{\mathrm{FM}}$,
while $\delta\mathbf{m}_1^{-}$ of $\Omega_{\mathrm{I}}$ [red dashed curve in Figs. \ref{fig5}(d)] 
always keeps right-handed polarization for $h\in(0,h^{\mathrm{FM}})$.
Alternatively, $\delta\mathbf{m}_1^{+}$ of $\Omega_{\mathrm{III}}$ [cyan solid curve in Figs. \ref{fig5}(c)] 
is always right-handed polarized for $0<h<h^{\mathrm{FM}}$. 
In particular, it tends to be linearly polarized in $\mathbf{e}_{z'}$ axis ($x'y$-plane)
when $h\rightarrow 0$ ($h\rightarrow h^{\mathrm{FM}}$).
As for $\delta\mathbf{m}_1^{-}$ of $\Omega_{\mathrm{III}}$ [cyan dashed curve in Figs. \ref{fig5}(d)],
it expands from linear polarization in $x'y$-plane around $h=0$ to right-handed and back to 
linear in $x'y$-plane at $h=h_{\mathrm{R}}$, then becomes left-handed for $h_{\mathrm{R}}<h<h^{\mathrm{FM}}$,
and finally becomes linearly polarized in $\mathbf{e}_{z'}$ axis when $h\rightarrow h^{\mathrm{FM}}$.
Across the whole region of $h\in(0,h^{\mathrm{FM}})$, the phase differences between $\delta\mathbf{m}_{1,2}^{+}$
for $\Omega_{\mathrm{I}}$ and $\Omega_{\mathrm{III}}$ remain $\pi$, 
while those  between $\delta\mathbf{m}_{1,2}^{-}$ are $0$.
Next we turn to the other hybridized pair: $\Omega_{\mathrm{II}}$ and $\Omega_{\mathrm{IV}}$.
If we take the following correspondence:
$\delta\mathbf{m}_1^{\pm}(\Omega_{\mathrm{I}})\leftrightarrow\delta\mathbf{m}_1^{\mp}(\Omega_{\mathrm{II}})$
and $\delta\mathbf{m}_1^{\pm}(\Omega_{\mathrm{III}})\leftrightarrow\delta\mathbf{m}_1^{\mp}(\Omega_{\mathrm{IV}})$,
the same behaviors described above hold. 
The only difference is that the left/right-hand transition point changes to $h_{\mathrm{L}}\approx 1.882$.
At last, the spin-wave intensity ratio, $|\delta\mathbf{m}_1^{\pm}|^2/|\delta\mathbf{m}_2^{\pm}|^2$, 
always equals to 1 for all frequency branches.

\begin{figure*} [htbp]
	\centering
	\includegraphics[width=1.0\textwidth]{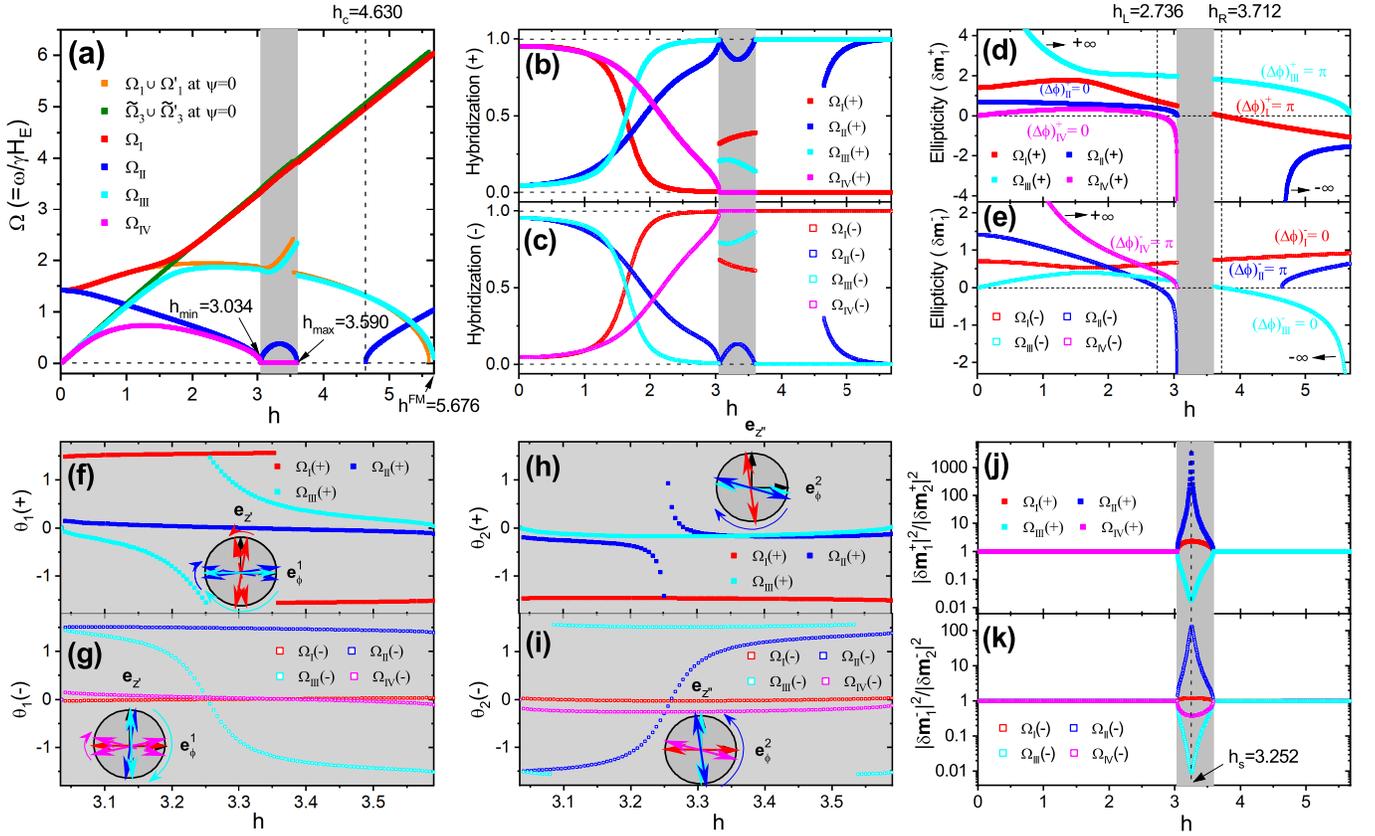}
	\caption{(Color online) Magnonics of symmetrical all-antiferromagnetic junctions
		with $k=1$ and $p=0.45$ under oblique dc fields with $\psi=\pi/10$. 
		The equilibrium magnetization layout falls into the SF phase when 
		$h\in \left(0,h_{\mathrm{min}}\right)\cup \left(h_{\mathrm{max}},h^{\mathrm{FM}}\right)$
		and MRS phase when $h\in \left[h_{\mathrm{min}},h_{\mathrm{max}}\right]$ 
		(shaded area in all subfigures).
		(a) Four positive eigenfrequencies $\Omega_{\mathrm{I}}$ (red), $\Omega_{\mathrm{II}}$ (blue), 
		$\Omega_{\mathrm{III}}$ (cyan),	and $\Omega_{\mathrm{IV}}$ (magenta). 
		The orange and olive curves are respectively $\Omega_{1}\cup\Omega'_{1}$ and
		 $\tilde{\Omega}_3\cup\tilde{\Omega}'_3$ for $\psi=0$ [see Fig. \ref{fig4}(a)].
		(b) and (c): Evolution of spin-wave intensities in subspaces with even and odd parities for all four eigenfrequencies. The parity operator for magnonics based on SF (MRS) phase
		is $\mathbb{C}$ ($\mathbb{M}_{xz}$).
		(d) and (e): Ellipticities of $\delta\mathbf{m}^{+}_{1}$ and $\delta\mathbf{m}^{-}_{1}$
		 for all four SF-phase-based eigenfrequencies, 
		 which are identical to those of $\delta\mathbf{m}^{\pm}_{2}$. 
		The corresponding phase differences between $\delta\mathbf{m}^{\pm}_{1,2}$ are indicated. 
		(f) and (g): Evolution of the linear polarization orientation of $\delta\mathbf{m}^{+}_{1}$
		and $\delta\mathbf{m}^{-}_{1}$ for $h\in \left[h_{\mathrm{min}},h_{\mathrm{max}}\right]$ 
		based on MRS phase. 
		(h) and (i): Counterparts of (f) and (g) for $\delta\mathbf{m}^{+}_{2}$
		and $\delta\mathbf{m}^{-}_{2}$.
		(j) and (k): Intensity ratio of the spin-wave component of group 1 over that of group 2
		in subspaces with even and odd parities under $\mathbb{C}$ ($\mathbb{M}_{xz}$) based on
		SF (MRS) phase.}\label{fig6}
\end{figure*}

When $p$ increases to $0.45$, the magnoncis becomes more complicated and the main features are plotted in Fig. \ref{fig6}.
Now the equilibrium magnetization layout falls in MRS phase when 
$h\in \left[h_{\mathrm{min}},h_{\mathrm{max}}\right]$ (shaded areas in all subfigures)
with $h_{\mathrm{min}}=3.034$ and $h_{\mathrm{max}}=3.590$, leading to quite a few novel behaviors of spin waves.
First, the four eigenfrequencies ($\Omega_{\mathrm{I\sim IV}}$) are provided in Fig. \ref{fig6}(a)
by red, blue, cyan and magenta solid curves in the order we have appointed.
Meantime the $\Omega_{1}\cup\Omega'_{1}$ and $\tilde{\Omega}_{3}\cup\tilde{\Omega}'_{3}$ branches at
$\psi=0$ have been appended by orange and olive curves, respectively.
We have not provided the $\tilde{\Omega}_{1}\cup\tilde{\Omega}'_{1}$ and $\Omega_{3}\cup\Omega'_{3}$ branches
in this subfigure since they are too close to $\Omega_{\mathrm{II,IV}}$.
Similar to ``$p<1/4$" case, the original two inter-parity crossings in SF phase are lifted by oblique dc fields
and become anticrossings.
To confirm this, the weights in subspaces with even and odd parities (under $\mathbb{C}$) for each
branch have been depicted in white areas of Figs. \ref{fig6}(b) and (c), respectively.
Clearly, for $\Omega_{\mathrm{I,IV}}$ ($\Omega_{\mathrm{II,III}}$)
 $|\delta\mathbf{m}_1^{+}|^2+|\delta\mathbf{m}_2^{+}|^2$
($|\delta\mathbf{m}_1^{-}|^2+|\delta\mathbf{m}_2^{-}|^2$) decreases from almost 1 to 0, indicating
the hybridization of subspaces with opposite parities under $\mathbb{C}$.
Spin waves in SF phase are all elliptically polarized and the ellipticities 
of $\delta\mathbf{m}_1^{\pm}$ (equal to those of $\delta\mathbf{m}_2^{\pm}$) are respectively
provided in Figs. \ref{fig6}(d) and (e), with positive (negative) values indicating 
right-handed (left-handed) polarization.
Except for the MRS-phase-based shaded area, the details are quite similar to those in 
Figs. \ref{fig5}(c) and (d) of $\psi=0$ case, so we will not go into it again.
The only difference is that $h_{\mathrm{L,R}}$ change to $h_{\mathrm{L}}=2.736$ and $h_{\mathrm{R}}=3.712$.
In addition, the spin-wave intensity ratio always equals to 1 for all branches in SF phase,
as shown in Figs. \ref{fig6}(j) and (k).

For spin waves in MRS phase, the symmetry operator changes to $\mathbb{M}_{xz}$.
Similar to in-plane-dc-field case, the $\pi/2$ phase difference between 
$\delta\mathbf{m}_1^{+}+\delta\mathbf{m}_2^{+}$ and $\delta\mathbf{m}_1^{-}+\delta\mathbf{m}_2^{-}$ still exists.
Except for this, the behaviors of spin waves become more fantastic.
First, the weight distributions in the even and odd subspaces under $\mathbb{M}_{xz}$ are respectively
plotted in the shaded areas in Figs. \ref{fig6}(b) and (c).
For $\Omega_{\mathrm{I,II,III}}$, the spin wave spans the two subspaces. 
While for $\Omega_{\mathrm{IV}}=0$, only subspace with odd parity accommodate the spin wave.
The polarization of all spin-wave components ($\delta\mathbf{m}_{1,2}^{\pm}$) become linear. 
Figures \ref{fig6}(f) and (g) show the polarization directions for $\delta\mathbf{m}_{1}^{\pm}$
in the local coordinate system $(\mathbf{e}_{\phi}^1,\mathbf{e}_{z'})$,
which differ from those of $\delta\mathbf{m}_{2}^{\pm}$ within $(\mathbf{e}_{\phi}^2,\mathbf{e}_{z^{\prime\prime}})$ 
presented in Figs. \ref{fig6}(h) and (i).
Specially, as $h$ increases from $h_{\mathrm{min}}$ to $h_{\mathrm{max}}$,
$\delta\mathbf{m}_{1}^{+}$ belonging to $\Omega_{\mathrm{I}}$, $\Omega_{\mathrm{II}}$
or $\Omega_{\mathrm{III}}$ respectively swings around $\mathbf{e}_{z'}$, $\mathbf{e}_{\phi}^1$ or
rotates clockwise from $+\mathbf{e}_{\phi}^1$ to $-\mathbf{e}_{\phi}^1$, as illustrated in Fig. \ref{fig6}(f).
In contrast, Fig. \ref{fig6}(h) shows that $\delta\mathbf{m}_{2}^{+}$ from $\Omega_{\mathrm{I}}$, 
$\Omega_{\mathrm{III}}$ or $\Omega_{\mathrm{II}}$ respectively swings around $\mathbf{e}_{z^{\prime\prime}}$,
$\mathbf{e}_{\phi}^2$ or rotates clockwise from $+\mathbf{e}_{\phi}^2$ to $-\mathbf{e}_{\phi}^2$.
Their intensity ratios, $|\delta\mathbf{m}_1^{+}|^2/|\delta\mathbf{m}_2^{+}|^2$, are plotted in 
the shaded area in Fig. \ref{fig6}(j). 
Clearly, they all deviate from 1.
In particular, $\delta\mathbf{m}_2^{+}$ ($\delta\mathbf{m}_1^{+}$) from $\Omega_{\mathrm{II}}$
($\Omega_{\mathrm{III}}$) disappears at $h_{\mathrm{s}}=3.252$.
Then we move to the subspace with odd parity (under $\mathbb{M}_{xz}$).
Now $\delta\mathbf{m}_{1}^{-}$ belonging to $\Omega_{\mathrm{I}}$, $\Omega_{\mathrm{IV}}$, $\Omega_{\mathrm{II}}$
or $\Omega_{\mathrm{III}}$ respectively swings around $\mathbf{e}_{\phi}^1$, $\mathbf{e}_{\phi}^1$, $\mathbf{e}_{z'}$, or
rotates clockwise from $+\mathbf{e}_{z'}$ to $-\mathbf{e}_{z'}$, as shown in Fig. \ref{fig6}(g).
Meantime, Fig. \ref{fig6}(i) provides that 
$\delta\mathbf{m}_{2}^{-}$ belonging to $\Omega_{\mathrm{I}}$, $\Omega_{\mathrm{IV}}$, $\Omega_{\mathrm{III}}$
or $\Omega_{\mathrm{II}}$ respectively swings around $\mathbf{e}_{\phi}^2$, $\mathbf{e}_{\phi}^2$,
$\mathbf{e}_{z^{\prime\prime}}$, or rotates anticlockwise from $-\mathbf{e}_{z^{\prime\prime}}$ 
to $+\mathbf{e}_{z^{\prime\prime}}$.
At last, $|\delta\mathbf{m}_1^{-}|^2/|\delta\mathbf{m}_2^{-}|^2$ for all four branches are provided in Fig. \ref{fig6}(k). 
Again, $\delta\mathbf{m}_2^{-}$ ($\delta\mathbf{m}_1^{-}$) from $\Omega_{\mathrm{II}}$
($\Omega_{\mathrm{III}}$) vanishes at $h_{\mathrm{s}}=3.252$.

\section{VI. Discussions and conclusion} 
First of all, in this work the in-plane anisotropy has been neglected. 
This is applicable for two reasons. On one hand, in real symmetric all-antiferromagnetic junctions
such as $\mathrm{Fe}_2\mathrm{O}_3$/$\mathrm{Cr}_2\mathrm{O}_3$/$\mathrm{Fe}_2\mathrm{O}_3$, the SF field
of the spacers is much higher than that of the outermost $\mathrm{Fe}_2\mathrm{O}_3$ sublayers.
Therefore the neglect of in-plane anisotropy in $\mathrm{Fe}_2\mathrm{O}_3$ sublayers is acceptable.
On the other hand, the ``easy-plane" anistropy of $\mathrm{Fe}_2\mathrm{O}_3$ sublayers 
reduces the concomitant SF field $H_{SF}\approx \sqrt{2H_E H_A}$ ($H_A$ being the in-plane anisotropy field) to zero,
thus greatly simplifies our analytics and does not lose the most significant features of these junctions.
Specifically, if we consider the in-plane anisotropy, the strict ``cruciferae" state at $h=0$ fails
and the ground state should exhibit some kind of hysteresis behavior.
SF and MRS phases will both survive, 
but the critical $p$ should be larger than $1/4$ from the ``easy-plane" case.
The reason why we did not explore further is that the central vectorial equality like
Eq. (\ref{Equality}), which is the basis of our analytics, is hard to obtain. 
Hence the magnetization statics and corresponding magnonics can hardly be analyzed theoretically. 
However, the main features we have proposed in the main text should persist.

Second, the magnonics in SF phase is relatively simple since the underlying symmetry operator
is the polarization-preserved rotation ($\mathbb{C}_{2x}$ or $\mathbb{C}$).
Spin waves therein are all elliptically polarized meantime bear fixed phase difference between components 
both from subspaces with opposite parities and from the same subspace but different group (1 or 2).
On the other hand, magnonics in MRS phase is more interesting since now the underlying symmetry 
operator becomes the polarization-broken mirror reflection $\mathbb{M}_{xz}$.
The most exciting feature is that oblique dc fields turn spin waves from elliptical to linear polarization.
This can be understood by analogy to the superposition of two circularly polarized light with 
the same frequency but opposite rotation direction into linearly polarized light.
In addition, the polarization direction of different components ($\delta\mathbf{m}_{1,2}^{\pm}$) from 
various frequency branches ($\Omega_{\mathrm{I\sim IV}}$) experience distinct swinging or rotating process
in the transverse plane (of $\mathbf{m}_{A,1(2)}^{\mathrm{eq}}$) as the oblique dc field increases within the MRS phase.
This provide a novel route of generating and manipulating linearly-polarized spin waves in
symmetric all-antiferromagnetic junctions.

Third, the emergence of MRS phase relies on a relatively large inter-sublayer orthogonal coupling
(in easy-plane case, $p=H_p/H_E>1/4$). In real all-antiferromagnetic junctions, this can be achieved
by appropriately decreasing the spacer thickness or fine-tuning the non-uniform domain wall states
in the spacers.
In addition, the two outermost antiferromagnetic sublayers should be symmetric.
If they are made of the same material but with different thickness, or just made of different materials,
the MRS phase should fade out since generally $\mathbf{m}_{A(B),1(2)}^{\mathrm{eq}}$ can not hold in the same
latitude circle. 
$\mathbb{M}_{xz}$ is no longer the strict symmetry operator then the combined spin waves 
change from linear to elliptical polarization (although can be very narrow).
Therefore to efficiently generate and manipulate linearly-polarized spin waves, the all-antiferromagnetic junctions
should be precisely prepared as a symmetric configuration.
Further investigations about the effects of asymmetry on magnetization statics and magnoncis
should be also interesting but beyond the scope of this work.

In summary, by appropriately reformulating the orthogonal coupling in Eq. (\ref{Coupling_energy_density}),
in this work the equilibrium magnetization layout and the corresponding coherent magnonics in 
symmetric all-antiferromagnetic junctions are systematically revealed.
Our contribution is mainly focused on three aspects:
(i) Under strong enough orthogonal coupling the equilibrium magnetization layout can fall into the MRS phase 
which is quite different from the usual SF phase that antiferromagnets often experience.
(ii) For in-plane dc fields, two inter-parity and two intra-parity accidental crossings exist in
the eigenfrequency spectrum of spin waves in SF phase. 
At $h=h_{\mathrm{min}}$ ($h_{\mathrm{max}}$)
where a second-order (first-order) PT occurs between SF and MRS phases, eigenfrequencies 
are continuous (discontinuous). Except for some isolated special points 
($h_{\mathrm{min}}$, $h_{\mathrm{max}}$ and $h_{\mathrm{c}}$),
spin waves are always elliptically polarized for $h\in(0,h^{\mathrm{FM}})$.
(iii) Oblique dc fields turn the inter-parity crossings in SF phase into anticrossings
meantime induce considerable hybridization between elliptically polarized spin wave components
belonging to subspaces with even and odd parities. 
The most exciting progress resides in magnonics based on MRS phase which is defined upon 
$\mathbb{M}_{xz}$. The spin wave components become linear polarized and the polarization
direction can be fine controlled by field strength within MRS phase. 
Our results lay the foundation for magnetic quantum phases and the corresponding coherent magnonics
in symmetric all-antiferromagnetic junctions, meantime open a new avenue for 
magnetic nanodevices with ultra-high density and ultra-fine control of magnonics.

\section{acknowledgments} 
M.L. acknowledges supports from the National Natural Science Foundation of China (Grant No. 12204403).
B.X. is funded by the National Natural Science Foundation of China (Grant No. 11774300).
W.H. is supported by the National Natural Science Foundation of China (Grant No. 12174427).

\appendix

\section{Appendix A: Explicit form of the quartic equation in Sec. IV.C}
\setcounter{equation}{0}
\renewcommand{\theequation}{A\arabic{equation}}

The parameters $J_{0,1,2,3}$ in the quartic algebraic equation of $\Omega^2$:
$(\Omega^2)^4 + J_3(\Omega^2)^3 +  J_2(\Omega^2)^2 +  J_1(\Omega^2) + J_0 =0$
has the following explicit form:
\begin{equation}\label{J3210_definition}
	\begin{split}
		J_0=&\left[\left(e_1\right)^2-\left(e_2\right)^2\right]\left[\left(f_1\right)^2-\left(f_2\right)^2\right], \\
		J_1=& 2\left(e_2 f_2 - e_1 f_1\right)\left[\left(a_1+a_2\right)\left(b_2+c_2\right)+\left(a_1-a_2\right)\left(b_1-c_1\right)\right],  \\
		        &+2\left(e_2 f_1 - e_1 f_2\right)\left[\left(a_1+a_2\right)\left(b_1-c_1\right)+\left(a_1-a_2\right)\left(b_2+c_2\right)\right],  \\
		        &+2e_1\left[\left(f_1\right)^2-\left(f_2\right)^2\right]+2f_1\left[\left(e_1\right)^2-\left(e_2\right)^2\right], \\
		J_2=&\left[\left(e_1+f_1\right)-\left(a_1+a_2\right)\left(b_2+c_2\right)-\left(a_1-a_2\right)\left(b_1-c_1\right)\right]^2 ,  \\
		        & -\left[\left(e_2-f_2\right)-\left(a_1+a_2\right)\left(b_1-c_1\right)-\left(a_1-a_2\right)\left(b_2+c_2\right)\right]^2 ,  \\
		        & +2\left(e_1f_1 - e_2 f_2\right), \\
		J_3=& 2\left[\left(e_1+f_1\right)-\left(a_1+a_2\right)\left(b_2+c_2\right)-\left(a_1-a_2\right)\left(b_1-c_1\right)\right],
	\end{split}
\end{equation}
with
\begin{equation}\label{a1_to_f2_definition}
	\begin{split}
		a_1=&1-4p\cos^2\epsilon, \\
		a_2=&1+4p\cos^2\epsilon,  \\
		b_1=&1+2p\sin^2 2\epsilon - 4p\cos^2\epsilon\cos2\epsilon + k\sin^2\epsilon\sin^2\chi,  \\
		b_2=& \cos2\epsilon-2p\sin^2 2\epsilon - 4p\cos^2\epsilon,  \\
		c_1=&1+ 4p\cos^2\epsilon\cos2\epsilon + k\sin^2\epsilon\sin^2\chi,  \\
        c_2=& \cos2\epsilon+ 4p\cos^2\epsilon,  \\
        d   =&\frac{k}{2}\sin\epsilon\sin2\chi,  \\
        e_1=&a_2\left(b_2-b_1\right)-kb_1\cos^2\chi+d^2,  \\
        e_2=&a_2\left(b_1-b_2\right)-kb_2\cos^2\chi \\
        f_1=&a_1\left(c_2-c_1\right)-kc_1\cos^2\chi+d^2,  \\
        f_2=&a_1\left(c_1-c_2\right)-kc_2\cos^2\chi.
	\end{split}
\end{equation}

\section{Appendix B: Detailed $P_{\mathrm{asym}}$ for MRS phase under oblique dc fields}
\setcounter{equation}{0}
\renewcommand{\theequation}{B\arabic{equation}}
After defining $\tilde{p}\equiv p\left[\left(\mathbf{m}_{A,1}^{\mathrm{eq}}+\mathbf{m}_{A,2}^{\mathrm{eq}}\right)\left(\mathbf{m}_{B,1}^{\mathrm{eq}}+\mathbf{m}_{B,2}^{\mathrm{eq}}\right)\right]=p\left\{\sin^2\theta_1\left[\left(\cos\phi_1+\cos\phi_2\right)^2-\left(\sin\phi_1-\sin\phi_2\right)^2\right]+4\cos^2\theta_1\right\}$,
$\mu\equiv\sin\eta\cos\epsilon-\sin\theta_1\cos\phi_2\cos\chi(1+\tan\chi\tan\xi)\sin\epsilon-\sin2\epsilon$,
$\nu\equiv\sin2\eta+\sin\theta_1\cos\phi_1\cos\xi(1+\tan\chi\tan\xi)\sin\eta-\sin\epsilon\cos\eta$,
and $\kappa\equiv\sin\theta_1(\cos\phi_2-\cos\phi_1)$,
the elements of $\tilde{\varTheta}$ and $\tilde{\varPhi}$ from $P_{\mathrm{asym}}$
in Eq. (\ref{OutOfPlaneDC_8x8_general}) are as follows:
\begin{equation}\label{varTheta_16_elements}
	\begin{split}
		\tilde{\varTheta}_{11}=&-\mathrm{i}k\sin\epsilon\sin\chi\cos\chi, \\
		\tilde{\varTheta}_{12}=&-\mathrm{i}(\tilde{p}-1)\sin\eta\sin(\xi-\chi),  \\
		\tilde{\varTheta}_{13}=&1-\tilde{p}+k\cos^2\chi,  \\
		\tilde{\varTheta}_{14}=&(1-\tilde{p})\cos(\xi-\chi),  \\
		\tilde{\varTheta}_{21}=&-\mathrm{i}(\tilde{p}-1)\sin\epsilon\sin(\xi-\chi),  \\
		\tilde{\varTheta}_{22}=&\mathrm{i}k\sin\eta\sin\xi\cos\xi, \\
		\tilde{\varTheta}_{23}=&\tilde{\varTheta}_{14}, \\
		\tilde{\varTheta}_{24}=&1-\tilde{p}+k\cos^2\xi,  \\
		\tilde{\varTheta}_{31}=&\tilde{p}\cos2\epsilon+1+k\sin^2\epsilon\sin^2\chi,  \\
		\tilde{\varTheta}_{32}=&(\tilde{p}+1)\cos\epsilon\cos\eta  \\
 		                          & \quad +(\tilde{p}-1)\sin\epsilon\sin\eta\cos(\xi-\chi),  \\
 		\tilde{\varTheta}_{33}=&-\tilde{\varTheta}_{11},   \\
 		\tilde{\varTheta}_{34}=&-\tilde{\varTheta}_{21},   \\
 		\tilde{\varTheta}_{41}=&\tilde{\varTheta}_{32},   \\
 		\tilde{\varTheta}_{42}=&\tilde{p}\cos2\eta+1+k\sin^2\eta\sin^2\xi,  \\
 		\tilde{\varTheta}_{43}=&-\tilde{\varTheta}_{12},  \\
 		\tilde{\varTheta}_{44}=&-\tilde{\varTheta}_{22},
	\end{split}
\end{equation}
and
\begin{equation}\label{varPhi_16_elements}
	\begin{split}
		\tilde{\varPhi}_{11}=& -\mathrm{i} k\sin\epsilon\sin\chi\cos\chi - 2\mathrm{i}p\mu\kappa\sin\chi, \\
		\tilde{\varPhi}_{12}=& \mathrm{i}(\tilde{p}+1)\sin\eta\sin(\xi-\chi)-2\mathrm{i}p\nu\kappa\sin\chi,   \\
		\tilde{\varPhi}_{13}=&(\tilde{p}+1+k\cos^2\chi)+2p\kappa^2\sin^2\chi,  \\
		\tilde{\varPhi}_{14}=&(\tilde{p}+1)\cos(\xi-\chi)-2p\kappa^2\sin\chi\sin\xi,  \\
		\tilde{\varPhi}_{21}=&\mathrm{i}(\tilde{p}+1)\sin\epsilon\sin(\xi-\chi)+2\mathrm{i}p\mu\kappa\sin\xi,  \\
		\tilde{\varPhi}_{22}=& \mathrm{i}k\sin\eta\sin\xi\cos\xi+2\mathrm{i}p\kappa\nu\sin\xi, \\
		\tilde{\varPhi}_{23}=&\tilde{\varPhi}_{14}, \\
		\tilde{\varPhi}_{24}=&(\tilde{p}+1+k\cos^2\xi)+2p\kappa^2\sin^2\xi,  \\
		\tilde{\varPhi}_{31}=&(1+k\sin^2\epsilon\sin^2\chi-\tilde{p}\cos2\epsilon)+2p\mu^2,  \\
		\tilde{\varPhi}_{32}=&2p\mu\nu+(1-\tilde{p})\cos\epsilon\cos\eta  \\
		                      & \quad -(1+\tilde{p})\sin\epsilon\sin\eta\cos(\xi-\chi),  \\
		\tilde{\varPhi}_{33}=&-\tilde{\varPhi}_{11},   \\
		\tilde{\varPhi}_{34}=&-\tilde{\varPhi}_{21},   \\
		\tilde{\varPhi}_{41}=&\tilde{\varPhi}_{32},   \\
		\tilde{\varPhi}_{42}=&(1+k\sin^2\eta\sin^2\xi-\tilde{p}\cos2\eta)+2p\nu^2,  \\
		\tilde{\varPhi}_{43}=&-\tilde{\varPhi}_{12},  \\
		\tilde{\varPhi}_{44}=&-\tilde{\varPhi}_{22}.
	\end{split}
\end{equation}


\end{document}